\documentclass[a4paper,fleqn]{cas-dc}

\usepackage[utf8]{inputenc}
\usepackage[numbers, sort&compress]{natbib}
\let\printorcid\relax 

\usepackage{tipa}

\usepackage{lineno,hyperref}
\usepackage{comment}
\usepackage{amsmath}
\usepackage{upgreek}
\usepackage{tikz}
\usepackage{gensymb}
\usepackage{soul}
\usepackage{multirow}
\usepackage[short]{optidef}
\usepackage{graphicx}
\usepackage{subfigure}
\usepackage{color}
\usepackage{graphics}
\usepackage{float}
\usepackage{tablefootnote}
\usepackage{tabularx}
\usepackage{algorithm,algorithmic}
\usepackage{subfigure}
\usepackage{booktabs}
\usepackage{framed} 
\usepackage{multicol} 

\usepackage{nomencl} 
\makenomenclature
\renewcommand*\nompreamble{\begin{multicols}{2}}
\renewcommand*\nompostamble{\end{multicols}}

\usepackage{ifthen}

\renewcommand{\nomgroup}[1]{%
	\ifthenelse{\equal{#1}{C}}{\item[\textit{Constants}]}
	{%
		\ifthenelse{\equal{#1}{V}}{\item[\textit{Variables}]}
		{%
			\ifthenelse{\equal{#1}{I}}{\item[\textit{Indices and sets}]}
			{%
				\ifthenelse{\equal{#1}{A}}{\item[\textit{Abbreviations}]}{}}}}
}

\def\tsc#1{\csdef{#1}{\textsc{\lowercase{#1}}\xspace}}
\tsc{WGM}
\tsc{QE}
\graphicspath{{./figures/}}

\definecolor{rev}{RGB}{0, 0, 0}
\newcommand{\rev}[1]{\textcolor{rev}{#1}}
\definecolor{revst}{RGB}{0, 0, 0}
\newcommand{\revst}[1]{\textcolor{revst}{#1}}

\begin{document}

\let\WriteBookmarks\relax
\def\floatpagepagefraction{1}
\def\textpagefraction{.001}

\shorttitle{Real-time Hybrid Controls of Energy Storage and Load Shedding for Integrated Power and Energy Systems of Ships}

\shortauthors{L. Vu et al. }

\title[mode = title]{Real-time Hybrid Controls of Energy Storage and Load Shedding for Integrated Power and Energy Systems of Ships}



\author[1]{Linh Vu}
\ead{litvu@clarkson.edu}

\author[2]{Thai-Thanh Nguyen}
\ead{thaithanh.nguyen@nypa.gov}

\author[3]{Bang Le-Huy Nguyen}
\ead{bangnguyen@ieee.org }

\author[1]{Md Isfakul Anam}
\ead{isfakum@clarson.edu}

\author[1]{Tuyen Vu}
\cormark[1]
\ead{tvu@clarkson.edu}
\cortext[cor1]{Corresponding author}

\address[1]{Clarkson University, Potsdam, NY, USA}
\address[2]{Advanced Grid Innovation Laboratory for Energy (AGILe), New York Power Authority (NYPA), NY, USA}
\address[3]{Los Alamos National Laboratory, Los Alamos, NM, USA}



\begin{abstract}
	This paper presents an original energy management methodology to enhance the resilience of ship power systems. The integration of various energy storage systems (ESS), including battery energy storage systems (BESS) and super-capacitor energy storage systems (SCESS), in modern ship power systems poses challenges in designing an efficient energy management system (EMS). The EMS proposed in this paper aims to achieve multiple objectives. The primary objective is to minimize shed loads, while the secondary objective is to effectively manage different types of ESS. Considering the diverse ramp-rate characteristics of generators, SCESS, and BESS, the proposed EMS exploits these differences to determine an optimal long-term schedule for minimizing shed loads. Furthermore, the proposed EMS balances the state-of-charge (SoC) of ESS and prioritizes the SCESS's SoC levels to ensure the efficient operation of BESS and SCESS. For better computational efficiency, we introduce the receding horizon optimization method, enabling real-time EMS implementation. A comparison with the fixed horizon optimization (FHO) validates its effectiveness. Simulation studies and results demonstrate that the proposed EMS efficiently manages generators, BESS, and SCESS, ensuring system resilience under generation shortages. Additionally, the proposed methodology significantly reduces the computational burden compared to the FHO technique while maintaining acceptable resilience performance.
\end{abstract}





\maketitle

\begin{table*}[!t]
	\begin{framed}
		\nomenclature[A]{ACLC}{\revst{AC load center}}
		\nomenclature[A]{AES}{All-electric ship}
		\nomenclature[A]{Aux PGM}{\revst{Auxiliary power generation module}}
		\nomenclature[A]{ESS}{Energy storage system}
		\nomenclature[A]{EMS}{Energy management system}
		\nomenclature[A]{GD}{\revst{Gradient descent}}
		\nomenclature[A]{BESS}{Battery energy storage system}
		\nomenclature[A]{SCESS}{Super-capacitor energy storage system}
		\nomenclature[A]{RHO}{Receding horizon optimization}
		\nomenclature[A]{FHO}{Fixed horizon optimization}
		\nomenclature[A]{SoC}{State-of-charge}
		\nomenclature[A]{HRRL}{High ramp-rate load}
		\nomenclature[A]{PCM}{Power conversion module}
		\nomenclature[A]{IPNC}{Integrated power node center}
		\nomenclature[A]{Main PGM}{\revst{Main power generation module}}
		\nomenclature[A]{PMM}{Propulsion motor module}
		\nomenclature[A]{ESM}{Energy storage module}
		\nomenclature[A]{SPS}{Ship power system}
		\nomenclature[A]{SWBD}{\revst{Switchboard}}
		\nomenclature[A]{MVAC}{\revst{Medium voltage alternating current }}
		\nomenclature[A]{MILP}{\revst{Mixed-integer linear program}}
		\nomenclature[C]{$n_L$}{Number of loads}
		\nomenclature[C]{$n_E$}{Number of ESS}
		\nomenclature[C]{$n_G$}{Number of generators}
		\nomenclature[C]{$N_p$}{Optimization horizon length}
		\nomenclature[C]{$T$}{Mission time}
		\nomenclature[C]{$\hat{w}_i$}{Load weight}
		\nomenclature[C]{$\alpha_e$}{Weight of SoC}

		\nomenclature[V]{$o_i^t$}{Operation status of load}
		\nomenclature[V]{$P_e^{E,t}$}{ESS power}
		\nomenclature[V]{$P_g^{G,t}$}{Generator power}
		\nomenclature[V]{$u_P^t$}{Auxiliary variable of absolute ESS power}
		\nomenclature[V]{$u_{SoC}^t$}{Auxiliary variable of SoC difference}
		\nomenclature[V]{$r_e^{E,t}$}{Power ramp-rate of ESS}
		\nomenclature[V]{$r_g^{G,t}$}{Power ramp-rate of generator}

		\nomenclature[I]{$t$}{Time index}
		\nomenclature[I]{$i$}{Load index}
		\nomenclature[I]{$e$}{ESS index}
		\nomenclature[I]{$g$}{Generator index}
		\printnomenclature
	\end{framed}
\end{table*}

\section{Introduction}


\begin{table}[]
	\caption{\revst{Characteristic comparison between lithium-Ion batteries and supercapacitors}}
	\label{tab:cha_compare}
	\revst{
		\begin{tabular}{@{}lcc@{}}
			\toprule
			\revst{Characteristic}                                              & BESS                     & SCESS                      \\ \midrule
			Energy Density (Wh/kg)                                              & 100 - 240                & 1 - 5                      \\
			Power Density (W/kg)                                                & 1,000 - 3,000            & 10,000$+$                  \\
			Cycle Life                                                          & 500$+$                   & 100,000$+$                 \\
			Average Life                                                        & 5 - 10 years             & 10 - 15 years              \\
			Safe Charging Temperature                                           & 0\degree C - 40\degree C & -40\degree C - 65\degree C \\
			\begin{tabular}[c]{@{}l@{}}Charge/Discharge Time\end{tabular}       & 30 - 600 m               & 1 - 10 s                   \\
			\begin{tabular}[c]{@{}l@{}}Charge/Discharge Efficiency\end{tabular} & 70 - 85 \%               & 85 - 98 \%                 \\ \bottomrule
		\end{tabular}
	}
\end{table}

\noindent \rev{Contemporary ship power systems (SPS) are undergoing a transition towards the adoption of fully electric vessels known as all-electric ships (AES). These AES implement sophisticated technologies including electric propulsion, energy storage systems (ESS), power conversion, and intelligent management systems. The integration of electrification into SPS facilitates the utilization of ESS to enhance the optimization of fuel consumption \cite{LAN201526, 6603331, 7793272} and adeptly manage substantial power ramp-rate demands \cite{9658612}.} However, the absence of tie-line connections renders the SPS vulnerable to system failures, particularly generation tripping. This limitation can be effectively mitigated by deploying ESS as a backup source. These ESS devices are crucial in efficiently compensating for power shortages resulting from generation unit tripping \cite{GEERTSMA201730, WEN2016158}. By integrating ESS into the AES, ship power systems gain the resilience necessary to overcome such challenges, thereby ensuring dependable and continuous operation.

\revst{ESS integrated into ships can be categorized based on their technological characteristics, as described in Table~\ref{tab:cha_compare} \cite{PAN2021111048}}. The first category comprises ESS with high energy density but low power density, exemplified by battery energy storage systems (BESS) capable of sustaining prolonged operation. In contrast, the second category includes super-capacitor energy storage systems (SCESS) possessing high power density but lower energy density for shorter durations. The ship's power system experiences diverse loads, including service and large pulse-power loads. Leveraging both types of ESS offers the advantage of combining high power density with high energy density, thereby enhancing the overall efficiency of the ship's power systems. BESS is well-suited for handling service loads and moderate ramp-rate loads due to its high energy density, while SCESS excels at serving critical high ramp-rate/transient loads due to its high power density \revst{and less sensitive to temperature variations} \cite{8638571, 9224178}. The coordination between these two types of ESS can be optimized to enhance the system's resiliency \cite{7512717}, ensuring reliable and efficient power delivery for modern AES.

The utilization of hybrid ESS, comprising BESS and SCESS in AES, has been explored in multiple studies \cite{7829349,8816352, 8928501,Zhang_2020, 9089227, 9991142}. Fuzzy control techniques have been employed to manage the hybrid ESS configuration effectively. Depending on specific requirements, various signal processing techniques such as low-pass filters \cite{7829349, 8816352}, high-pass filters \cite{8928501}, and ensemble empirical mode decomposition \cite{Zhang_2020} were utilized to separate low and high-frequency components, generating reference signals for BESS and SCESS using fuzzy controls. In \cite{9991142}, a flatness control approach was employed to reduce the complexity of ship system models. This reduction facilitated updating reference signals at a higher frequency, which proved beneficial for accommodating the dynamic response of transient loads. However, these strategies might not be fully suitable for managing SPS during generation shortage conditions, as they were not explicitly designed to handle multiperiod scheduling of ESS. Therefore, further research and development of specialized control methodologies are necessary to address the unique challenges and complexities of such scenarios effectively.

Model predictive controls have been widely used for multiperiod scheduling of ESS in conventional power systems. In ship power systems, model predictive controls for hybrid ESS, as presented in \cite{HOU201862, HASELTALAB2019113308, 9206539}, have been utilized to mitigate load fluctuations and enhance fuel efficiency. Additionally, in \cite{7894212, 10220511}, model predictive control was employed to optimize the coordination between power ESS and generators during high-power ramp-rate conditions. However, these studies primarily focused on the optimal management of ESS under the assumption of sufficient generation capacity to supply loads, rather than addressing the resilient enhancement of the ship power systems. In situations where generation shortages occur, a different approach is necessary to ensure system stability and prevent widespread outages. Load shedding control becomes essential to shed noncritical loads, preserve critical loads, and sustain system functionality. To effectively manage generation, including hybrid ESS, and minimize shed loads, an energy management system (EMS) oriented towards resilience is required. This resilience-oriented EMS should take into account the unique ramp-rate characteristics of generators and ESS, playing a vital role in future SPS.

Resilience-oriented operation in ship power systems aims to selectively shed loads during periods of insufficient generation to preserve essential loads. Load importance is classified into three categories: vital, semivital, and nonvital, based on assigned weight values. Shedding nonvital loads with low weight values becomes a priority when power generation falls below the load demands. The overarching goal of enhancing ship power systems' resilience involves optimizing the scheduling of ESS to maximize load operability. Studies on resilience enhancement in SPS can be broadly categorized into centralized and decentralized approaches. In the centralized approach, a single control center tackles the optimization problem, while in the decentralized approach, multiple control units collaborate to solve the optimization task.

\begin{table*}[]
	\revst{
		\caption{Literature review of energy management systems for ship power systems}
		\label{tab:lit_comp}
		\begin{tabular}{@{}clcccccc@{}}
			\toprule
			Reference                                &
			Method                                   &
			\multicolumn{2}{c}{ESS inclusion}        &
			Case study size                          &
			Horizon length                           &
			Timestep                                 &
			Solution time                                                                                                                                \\ \cmidrule(lr){3-4}
			                                         & \multicolumn{1}{c}{}     & BESS       & SCESS      &          &        &              &           \\ \midrule
			\cite{6197250}                           & Multi-agent              & \checkmark &            & 40 MW    & -      & 5 $\mu$s     & -         \\
			\cite{6876221}                           & Multi-agent              & \checkmark &            & 40 MW    & -      & 10 ms        & -         \\
			\cite{7055266}                           & MPC                      &            &            & 103 kW   & 5      & 20 ms        & 1.5 ms    \\
			\cite{HOU2018919}                        & MPC                      & \checkmark &            & 3.5 kW   & 90,000 & 0.02 s       & -         \\
			\cite{MOUSAVIZADEH2018443}               & MILP                     & \checkmark &            & 22.71 MW & -      & 1 h          & 21 m      \\
			\cite{7894212}                           & MPC                      & \checkmark & \checkmark & 14 kW    & 500    & 10 ms        & -         \\
			\cite{7862235}                           & DCA                      & \checkmark &            & 8.8 kW   & -      & 100 ms       & -         \\
			\cite{LAI2018821}                        & ADMM                     & \checkmark &            & 375 kW   & -      & 30 m         & -         \\
			\cite{HOU201862}                         & MPC                      & \checkmark & \checkmark & 2 kW     & 20     & 0.01 s       & -         \\
			\cite{8466001}                           & LNBD                     & \checkmark &            & 14 MW    & -      & 1 h          & 198 s     \\
			\cite{8638571}                           & Two-step multi-objective & \checkmark & \checkmark & 30 MW    & -      & 48 m         & -         \\
			\cite{8918031}                           & MILP                     & \checkmark &            & 57.5 MW  & -      & 1 d          & 5.37 h    \\
			\cite{8945168}                           & MPC                      & \checkmark & \checkmark & 300 kW   & 1      & 100 $\mu$s   & -         \\
			\cite{doi:10.1080/20464177.2019.1684122} & DMPC                     & \checkmark &            & 140 MW   & 125    & 10 ms        & -         \\
			\cite{VAFAMAND2020550}                   & MPC                      & \checkmark &            & 765 kW   & 10     & 1 m          & -         \\
			\cite{9658612}                           & MPC                      & \checkmark &            & 59 MW    & 5      & 100 $\mu$s   & -         \\
			\cite{9645951}                           & MILP                     & \checkmark &            & 5.38 MW  & -      & 1 s          & -         \\
			\cite{9266062}                           & SCA                      & \checkmark &            & 49.5 MW  & -      & 30 m         & 84.95 s   \\
			\cite{9089227}                           & SCA                      & \checkmark &            & 600 kW   & -      & 24 h         & -         \\
			\cite{9224178}                           & Dynamic   programing     & \checkmark &            & 1 MW     & -      & 30 m         & -         \\
			\cite{jmse9090993}                       & MPC                      & \checkmark &            & 7.5 MW   & 2      & 20 s         & -         \\
			\cite{PLANAKIS2021104795}                & MPC                      & \checkmark &            & 351 kW   & 20     & 0.5 s        & 20 ms     \\
			\cite{9720160}                           & Multirate control        & \checkmark &            & 20 MW    & -      & 0.03 s       & -         \\
			\cite{ZHANG2022104763}                   & Two-level   MPC          & \checkmark & \checkmark & 30 kW    & 20 | 5 & 125 s | 25 s & -         \\
			\cite{CHEN2023109319}                    & MPC                      & \checkmark &            & -        & 10     & 1 s          & -         \\
			\cite{9921351}                           & Markov approximation     & \checkmark &            & 1.25 MW  & -      & 10 s         & -         \\
			\cite{10271302}                          & MPC                      & \checkmark &            & 30 MW    & -      & 0.1 s        & -         \\
			\cite{10226508}                          & RL, MPC                  & \checkmark &            & 610 kW   & 600    & 0.1 s        & -         \\
			\cite{LIU2024113894}                     & Hierarchical MPC         & \checkmark & \checkmark & 2.6 MW   & 5      & 0.1 s        & 119.35 ms \\ \bottomrule
		\end{tabular}}
\end{table*}

The centralized approaches offer several advantages, including broad supervision, straightforward implementation, and high accuracy in dispatches. Various centralized operation strategies and optimization methods have been presented to enhance the system resilience, such as probabilistic methods \cite{9131191, MOMOH2002145, 7055266}, a two-phase optimization problem \cite{8374982, 9244602}, a graph-theoretic method \cite{8423075}, dynamic prioritization approaches \cite{4957530, 4906545}, an adaptive risk-averse stochastic programming \cite{9266062}, and reconfiguration approach in \cite{8466001}. As these solutions did not take ESS into consideration, they are inappropriate for the modern ship power system with ESS integration. In addition, although ESS was not involved in \cite{8466001, 8374982, 8423075}, the optimization problems investigated within these studies remain complex and pose significant challenges in terms of their resolution. To address the complexity of the centralized optimization problem, constraint relaxation \cite{9645951} or a combination with rule-based methods \cite{https://doi.org/10.1049/iet-gtd.2020.0668} have been employed, resulting in a new low-complexity problem formulation that ensures feasible near-optimal solutions. Another approach to mitigate complexity is the adoption of multiperiod optimization, as presented in \cite{9207909}. However, the computational burden associated with the size and complexity of such problems over a long horizon significantly limits their real-time applicability. Notably, none of the aforementioned solutions have been evaluated specifically for real-time applications.

Distributed strategies in \cite{LAI2018821, doi:10.1080/20464177.2019.1684122, MOHAMED2020118041, 8847883, 6876221, 6197250, 9720160} overcome the computational limitation of the centralized approaches as multiple controllers solve the global optimization problem. In \cite{LAI2018821, doi:10.1080/20464177.2019.1684122, MOHAMED2020118041, 8847883}, \rev{ship power systems were divided into several zones, and the alternating direction method of multipliers algorithm were employed to address the EMS challenges in a decentralized manner. Multiagent strategies were introduced in \cite{6876221, 6197250} to tackle load management problems in SPS. Nevertheless, the distributed EMS methods outlined in these works are encumbered by challenges, including the intricacies of communication systems and potential cybersecurity vulnerabilities arising from their reliance on communication networks. Moreover, such decentralized strategies only attain limited accuracy \cite{8186925}.}

\rev{Existing energy management approaches for SPS have been extensively studied, and they exhibit both advantages and disadvantages. Centralized methods offer the benefits of simplicity and high accuracy in their energy management strategies. They are, nevertheless, computationally costly, especially when dealing with a high number of variables \cite{8641315, 8918031, 9224178}. Decentralized approaches, on the other hand, reduce the computing cost by tackling the problem in a distributed fashion. Nevertheless, these methods generally achieve only moderate accuracy and face challenges related to cyber-security concerns \cite{9026756, 10238789, 9205672}.} To address the above-mentioned problems, this paper proposes the use of receding horizon optimization (RHO) to optimize the coordination between hybrid ESS under insufficient generation conditions of ship power systems. Although RHO has been studied extensively in conventional power systems, its application in \rev{SPS} is limited due to the complexity of optimization problems, especially for real-time applications for ship power systems with the integration of hybrid ESS. To the best of the author's knowledge, this study is the first to address the coordination of hybrid ESS in the ship power system using RHO. The advantages of the proposed approach compared to existing centralized and decentralized strategies in the ship power systems are as follows:
\begin{itemize}
	\item The proposed methodology adopts a centralized approach; however, it offers a computational advantage compared to existing centralized methods. This advantage stems from the use of RHO, which solves the optimization problem using a sequence of short trajectories, as opposed to a single large trajectory in FHO.
	\item The proposed method exhibits a computational effort similar to that of distributed methods, but it holds an advantage in terms of ease of implementation. This advantage arises from its simplicity, as it does not require complex communication networks or synchronization mechanisms typically associated with distributed approaches.
\end{itemize}

In addition, existing studies have not considered the practical constraint of ESS, such as balancing state-of-charge (SoC) among multiple ESS. As the SoC estimation is not perfect, the long-term operation would cause significant SoC variances among ESS. The ESS with low SoC might cease to operate, causing overcurrent in other ESS and unintentional outages \cite{8945168, 7862235}. Furthermore, prioritizing a specific type of ESS for a particular role is an essential factor as different types of ESS have different characteristics. For instance, the SCESS is prioritized to serve high ramp-rate loads. However, such important issues have not been addressed in existing resilient-oriented methodologies. A multi-objective optimization problem is proposed in this paper to address these issues. In the proposed methodology, a primary objective is to minimize shed loads, and a secondary objective is to effectively manage different ESS types in ship power systems.

\textbf{This paper's main contributions are as follows:}
\begin{itemize}
	\item A resilience-oriented EMS taking advantage of different types of ESS into account is proposed. Compared to existing methods in \rev{\cite{9131191, MOMOH2002145, 7055266,8374982, 8423075, 4957530, 4906545, 8466001}}, in which the ESS was  disregarded, the proposed EMS is more applicable for modern ship power systems integrated with multiple types of ESS for multiple mission purposees.
	\item \rev{The proposed EMS considers practical and essential constraints of ESS, including
		      the equilibrium of SoC across ESS units and according priority the SoC level of SCESS to ensure the accomplishment of mission objectives.}
	\item The receding horizon optimization is used for enhancing the coordination of hybrid ESS and conventional generators within ship power systems. The methodology presented in this work not only tackles the computational challenges discussed in \cite{9207909}, but also addresses the communication network-related concerns highlighted in \cite{LAI2018821, doi:10.1080/20464177.2019.1684122, MOHAMED2020118041, 8847883, 6876221, 6197250}. Through this integrated approach, the proposed solution proves to be well-suited for real-time operational scenarios.

\end{itemize}

The rest of this paper is organized as follows: Section~\ref{sec:framework} presents an overall framework of the proposed approach. Section~\ref{sec:problem} describes a problem formulation to enhance the resilience of ship power systems. A multi-objective optimization problem is explained first, and then the detailed linearization process to solve the optimization problem by the mixed-integer linear program (MILP) is given. The use of receding horizon optimization for the proposed problem is presented in Section~\ref{sec:RHO}. Case studies of notional four-zone MVAC ship power systems are described in Section~\ref{sec: casestudies}. A comparison study on the RHO and FHO methods is also given in this section. \rev{In conclusion, the primary outcomes of this research are outlined in Section~\ref{sec:conclusion}.}

\section{Proposed Energy Management Methodology}
\label{sec:framework}

\begin{figure}[t]
	\centering
	\includegraphics[width=1\linewidth]{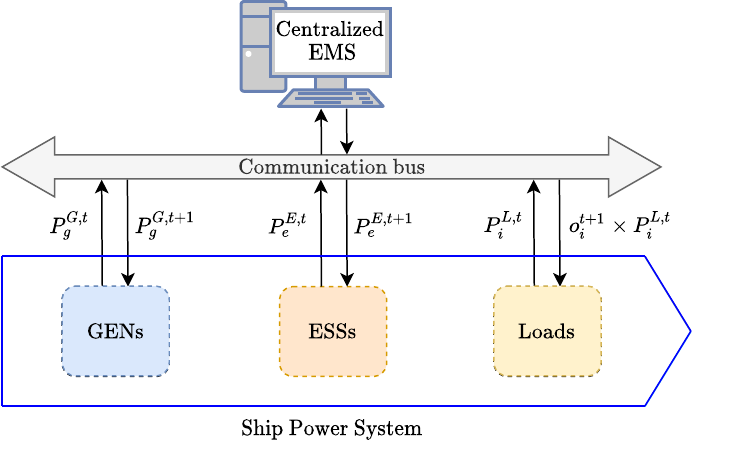}
	\caption{Overview of energy management system for ship power systems.}
	\label{fig:EMShip}
\end{figure}

\noindent Fig.~\ref{fig:EMShip} illustrates the EMS for SPS, encompassing communication connections. The central EMS collects comprehensive data concerning generators (GENs), ESS, and loads. Subsequently, an optimization procedure is executed to determine the operational condition of loads, along with the power levels of ESS and GENs. These optimized parameters are subsequently fed back to the SPS as reference signals. Localized controllers governing the GENs and ESS respond proactively to these reference inputs. In cases where the operational condition value falls below the commanded operational status, load shedding mechanisms are activated.

It should be noted that there is no standard of resilience metrics until now. However, several resilience metrics have been presented, which are based on the resilience features, reliability, etc. \cite{8966351}. The resistancy metric, which is defined in \cite{MOUSAVIZADEH2018443} as the ratio of served load to the total load demand considering the load priority factor, is used in this paper to evaluate the effectiveness of the proposed methodology.

The significance of loads is characterized by a weight parameter $w_i$, subject to adjustments based on mission operations, thereby causing the load's weight to be flexibly adapted across diverse missions. The quantification of load operability ($O$), as formulated in (\ref{eq:Oi}), gauges the extent to which loads are catered for. This formulation, adapted from \cite{5705696}, is modified to ensure the integration of load prioritization and corresponding rating power into load management problems. The enhancement of SPS's resilience is achieved through the optimal scheduling of ESS and generators to maximize the load operability metric.

\begin{align}
	\label{eq:Oi}
	O         & = {\frac{{\int_{t_0}^{t_f}}{\sum_{i = 1}^{n_L}}{\hat{w}_i o_i^t} dt
	}{
	{\int_{t_0}^{t_f}}{\sum_{i = 1}^{n_L}}{\hat{w}_i o_i^{*t}} dt}
	},                                                                                                   \\
	\hat{w}_i & = w_i P_{Li}^{\text{rated}}                                         & \forall i \in n_L,
\end{align}
\revst{where $n_L$ denotes the count of loads in the SPS; $P_{Li}^{\text{rated}}$ signifies the rated power for load $i$; $w_i$ corresponds to the weight value of load $i$; $\hat{w}_i$ corresponds to the normalized weight value pertaining to load $i$; $o_i^t$ signifies the operational state of load $i$ at the instance $t$; $o_i^{*t}$ is the commanded operational status of load $i$ at time $t$. The interval is described by $t_0$ as the event's commencement time and $t_f$ as its conclusion time. The load operability metric ($O$) quantifies the actual performance of the SPS under the commanded operational status. $O$ takes values in the range $[0, 1]$ due to system uncertainties, such as generator shortages and cyberattacks.}

\section{Problem Formulation}
\label{sec:problem}
\subsection{Objective Function}


\noindent The objective function of the proposed EMS is given by~(\ref{eq:objfunc}), which includes the four following terms:
\begin{itemize}
	\item The first term, $f_1(o_i^t)$ in ~(\ref{eq:objfuncf1}), is the total load operability over mission time $T$. By maximizing the total load operability, the amounts of shed loads are minimized, thus enhancing the system's resilience.
	\item The second term, $f_2(P_e^{E,t})$ in ~(\ref{eq:objfuncf2}), is a total of absolute ESS power over time $T$. As the ESS can charge and discharge, they can exchange power to balance their SoC levels. This operation is insufficient due to power losses of ESS during internal charging and discharging among ESS. This issue is mitigated by minimizing the total of absolute ESS power, as given by~(\ref{eq:objfuncf2}).
	\item The third term, $f_3(\text{SoC}^t)$, \rev{represents the divergence in SoC between ESS. The equilibrium of SoC levels across ESS is achieved by minimizing this term.}
	\item The final term, $f_4(\text{SoC}^{T})$, is the total of final SoC levels at the conclusion of the optimization window. Maximizing the final SoC levels ensures that all ESS have sufficient reserves for future load demand. This feature makes the proposed method distinguish from existing optimization algorithms, like in \cite{9207909}, that maintain the same SoC levels at initial and final intervals of the optimization window. \rev{Opting for an increased $\alpha_e$ value prioritizes the charging of SCESS to cater to high ramp-rate loads. The ultimate component of the objective function aids in enabling the receding horizon optimization to effectively sustain the SoC level over extended operational periods.}
\end{itemize}

\begin{maxi}[1]
	{x}
	{	f(x) = f_1(o_i^t)
	- \omega_1 f_2(P_e^{E,t})
	}
	{}{}
	\breakObjective{- \omega_2 f_3(\text{SoC}^t) + \omega_3 f_4(\text{SoC}^{T})},
	\label{eq:objfunc}
\end{maxi}
\begin{align}
	x & = [o_i^t,P_e^{E,t}, P_g^{G,t}]^\top,                                     \\
	\label{eq:objfuncf1}
	f_1(o_i^t)
	  & = {
	\sum_{t = 1}^T
	{\sum_{i = 1}^{n_L}{\hat{w}_i o_i^t}}},                                      \\
	f_2(P_e^{E,t})
	\label{eq:objfuncf2}
	  & = \sum_{t = 1}^T{\sum_{e = 1}^{n_E}{|P_e^{E,t}|}},                       \\
	f_3(\text{SoC}^t)
	\label{eq:objfuncf3}
	  & = \sum_{t = 1}^T{\sum_{l,m}{|\text{SoC}_l^{E,t} - \text{SoC}_m^{E,t}|}}, \\
	f_4(\text{SoC}^{T})
	\label{eq:objfuncf4}
	  & = \sum_{e = 1}^{n_E}{\alpha_e \text{SoC}_e^{E,T}},
\end{align}
\rev{where $T$ denotes the duration of the mission; $o_i^t$ stands for the operational status of load $i$ during time $t$; $\hat{w}_i$ represents the load weight, indicating its significance as critical or non-critical; $P_e^{E,t}$ signifies the active power of ESS $e$ at time $t$; $\text{SoC}^t$ portrays SoC of EES at time $t$; while $\text{SoC}^T$ characterizes the ESS's SoC at the conclusion of the optimization window. Additionally, the constants $\omega_1$, $\omega_2$, and $\omega_3$ are adjustable weights that will be optimized using the gradient descent technique, as detailed in Section \ref{sec: gradient decent}.}

\subsection{Constraints}

\noindent \rev{The optimization goal presented in (\ref{eq:objfunc}) is bound by the power balance constraint specified in (\ref{eq:dP}). This constraint ensures that the optimized load operability remains below the aggregate generation, encompassing ESS power.}
\revst{
	\begin{align}
		\label{eq:dP}
		\sum_{i=1}^{n_L}{P_{i}^{L,t}} o_i^t \leq &
		\sum_{e = 1}^{n_E}{P_e^{E,t}} +
		\beta\sum_{g = 1}^{n_G}{P_g^{G,t}}
		                                         & {\forall t \in T},
	\end{align}}
\rev{herein, $n_L$ signifies the number of loads; \revst{$\beta = 0.95$  corresponds to the spinning reserve of the generator;} $P_{i}^{L,t}$ represents the commanded power for load $i$ at time $t$; $n_E$ is the number of ESS; $n_G$ corresponds to the number of generators; and $P_g^G$ denotes the power output of generator $g$.}

\rev{
The operational status $o_i^t$ adheres to the boundary restrictions outlined in (\ref{eq:oilim}), where the prescribed operational status equals unity ($o_i^{*t} = 1$). When $o_i^t = 0$, load $i$ must be completely shed at time $t$; conversely, when $o_i^t = 1$, load $i$ is fully serviced at time $t$. Furthermore, the ESS and generators are also subject to corresponding boundary limitations, as detailed in equations (\ref{eq:Pelim}) and (\ref{eq:Pglim}).
}
\begin{align}
	\label{eq:oilim}
	0          & \leq o_i^t \leq 1              & {\forall i \in n_L, \forall t \in T}, \\
	\label{eq:Pelim}
	P_e^{\min} & \leq P_e^{E,t} \leq P_e^{\max} & {\forall e \in n_E, \forall t \in T}, \\
	\label{eq:Pglim}
	P_g^{\min} & \leq P_g^{G,t} \leq P_g^{\max} & {\forall g \in n_G, \forall t \in T},
\end{align}
\rev{herein, $P_e^{\min}$ and $P_e^{\max}$ denote the lower and upper limits of ESS capacity, respectively. Similarly, $P_g^{\max}$ and $P_g^{\min}$ represent the maximum and minimum capacities of generators, respectively.}

\rev{Given that certain loads exhibit discrete variations, the operational state of such loads is governed by a discrete function, outlined as (\ref{eq:oistep}).}

\begin{align}
	\label{eq:oistep}
	o_i^t \in
	\begin{cases}
		\mathcal{Z} & \text{if load } i \text{ is the discrete load,} \\
		[0, 1]      & \text{otherwise},
	\end{cases}
\end{align}
where $\mathcal{Z} = \{0:\Delta o_i:1\}$; $\Delta o_i = 1/n$; $n \in \mathbb{Z}^+$ is the number of steps (the granularity) of the load command.

\rev{Given that both EES and generators are constrained by power ramp-rate limitations, the designated power reference points for these components are bound by ramp-rate restrictions as detailed in (\ref{eq:rglim}) and (\ref{eq:relim}).}
\begin{align}
	\label{eq:rglim}
	r_g^{\min} & \leq r_g^{G,t} \leq r_g^{\max}               & {\forall g \in n_G, \forall t \in T}, \\
	\label{eq:relim}
	r_e^{\min} & \leq r_e^{E,t} \leq r_e^{\max}               & {\forall e \in n_E, \forall t \in T}, \\
	r_g^{G,t}  & = {\frac{P_g^{G,t} - P_g^{G,t-1}}{\Delta t}} & {\forall g \in n_G, \forall t \in T}, \\
	r_e^{E,t}  & = {\frac{P_e^{E,t} - P_e^{E,t-1}}{\Delta t}} & {\forall e \in n_E, \forall t \in T},
\end{align}
where $\Delta t$ is the sampling time.
\rev{Ultimately, in order to mitigate abrupt charging and discharging of the ESS, the SoC level is constrained within specified lower and upper} (\ref{eq:SoClim}).
\begin{flalign}
	\label{eq:SoClim}
	\text{SoC}_e^{\min} & \leq \text{SoC}_e^{E,t} \leq \text{SoC}_e^{\max}                 & { \forall e \in n_E, \forall t \in T}, &  & \\
	\text{SoC}_e^{E,t}  & = \text{SoC}_e^{E,t-1} + \Delta t \frac{{P_e^{E,t}}}{P_e^{\max}} & {\forall e \in n_E, \forall t \in T}.  &  &
\end{flalign}

\subsection{MILP to Solve Optimization Problem}

\noindent The problem formulated from Section \ref{sec:problem} includes discrete variables of operational status and nonlinear objective terms ($f_2$ and $f_3$). To solve (\ref{eq:objfunc}) by MILP, the discrete variables are converted to integer variables and the nonlinear objective terms are linearized. To convert discrete variables to integer variables, the load power ($P_{i}^{L,t}$) and load weight value ($\hat{w}_i$) of the discrete loads are scaled with a factor of step size ($\Delta o_i$), as given in (\ref{eq:scaledwi}) and (\ref{eq:scaledPL}).
\begin{align}
	\label{eq:scaledwi}
	\hat{w}_i         & = \hat{w}_i \Delta o_i     \\
	\label{eq:scaledPL}
	\hat{P}_{i}^{L,t} & = {P_{i}^{L,t} \Delta o_i}
\end{align}
where $\hat{w}_i$ and $\hat{P}_{i}^{L,t}$ are the modified load weight and power, respectively. Substituting (\ref{eq:scaledwi}) and (\ref{eq:scaledPL}) into (\ref{eq:objfuncf1}) and (\ref{eq:dP}), respectively, the variables of operational status are converted into the integer variables subjected to the following constraint.
\begin{align}
	\label{eq:limoihat}
	0 \leq o_i^t \leq {\frac{1}{\Delta o_i}}
\end{align}

The optimal variables of operation status found by the optimization process are converted back to discrete variables using (\ref{eq:scaledoi2}), which will be used as the load commands for the discrete loads.
\begin{align}
	\label{eq:scaledoi2}
	\hat{o}_i^t = { o_i^t \Delta o_i}
\end{align}

Nonlinear objective terms are linearized by introducing auxiliary variables and constraints for such variables, as given by (\ref{eq:aux_uP}) and (\ref{eq:aux_uSoC}).
\begin{align}
	\label{eq:aux_uP}
	u_P^t     & = |P_e^{E,t}|,                              \\
	\label{eq:aux_uSoC}
	u_{SoC}^t & = |\text{SoC}_l^{E,t} - \text{SoC}_m^{E,t}|
\end{align}

The objective function (\ref{eq:objfunc}) can be represented by (\ref{eq:objfunc_linearize}) subjected to new constraints of operational status, power balance, and additional constraints of auxiliary variables. Thus, MILP can be used to solve the problem (\ref{eq:objfunc_linearize}).
\begin{maxi}[1]
	{x}
	{	f(x) = f_1(o_i^t)
		- \omega_1 f_2(u_P^t)
	}
	{}{}
	\breakObjective{- \omega_2 f_3(u_{SoC}^t) + \omega_3 f_4(\text{SoC}^{T}),}
	\addConstraint{0& \leq o_i^t \leq {\frac{1}{\Delta o_i}}}{\forall i \in n_L, \forall t \in T}
	\addConstraint{\sum_{i=1}^{n_L}{\hat{P}_{i}^{L,t}} o_i^t & \leq \sum_{e = 1}^{n_{E,t}}{P_e^E} + \sum_{g = 1}^{n_G}{P_g^{G,t}}} {\forall t \in T}
	\addConstraint{0 & \leq u_P^t \leq P_e^{\max}}{ \forall e \in n_E, \forall t \in T}
	\addConstraint{u_P^t & \geq  P_e^{E,t}}{\forall e \in n_E, \forall t \in T}
	\addConstraint{u_P^t & \geq  -P_e^{E,t}}{\forall e \in n_E, \forall t \in T}
	\addConstraint{0 & \leq u_{SoC}^t \leq \text{SoC}_e^{\max}}{ \quad\forall e \in n_E, \forall t \in T}
	\addConstraint{u_{SoC}^t & \geq \text{SoC}_l^{E,t} - \text{SoC}_m^{E,t} }{\forall l, m \in n_E, \forall t \in T}
	\addConstraint{u_{SoC}^t & \geq -\text{SoC}_l^{E,t} + \text{SoC}_m^{E,t} }{\forall l, m \in n_E, \forall t \in T},
	\label{eq:objfunc_linearize}
\end{maxi}
where
\begin{align}
	x & = [o_i^t,P_e^{E,t}, P_g^{G,t}, u_P^t, u_{SoC}^t]^\top, \\
	\label{eq:objfuncf12}
	f_1(o_i^t)
	  & = {
	\sum_{t = 1}^T
	{\sum_{i = 1}^{n_L}{\hat{w}_i o_i^t}}},                    \\
	f_2(u_P^t)
	\label{eq:objfuncf2u}
	  & = \sum_{t = 1}^T{\sum_{e = 1}^{n_E}{u_P^t}},           \\
	f_3(u_{SoC}^t)
	\label{eq:objfuncf3u}
	  & = \sum_{t = 1}^T{\sum_{l,m}{u_{SoC}^t}}.
\end{align}

\section{Proposed Receding Horizon Optimization}
\label{sec:RHO}

\noindent \rev{The problem fomulated in (\ref{eq:objfunc}) can be addressed using the FHO technique, where the entire mission duration forms the optimization window. Solving the problem detailed in (\ref{eq:objfunc}) provides the optimal solution for all time intervals. However, the robust functioning of the SPS necessitates a finer control time step (0.5~s) to accommodate high ramp-rate loads. This need for a smaller time increment results in a substantial expansion of the optimization problem's size. Moreover, the adoption of MILP poses additional complexities due to its inherent nonlinearity, significantly impacting the computational demands inherent in real-time implementation.}

\rev{To address the aforementioned challenge, this paper adopts the receding horizon optimization (RHO) approach. The RHO's framework is structured around a sequence of shorter time intervals, diverging from the extended trajectory of the FHO method. In RHO, the optimization problem is tackled at each discrete time step, resulting in a set of solutions spanning a fixed time window. Specifically, only the solution for the first time step within each window is executed on the ship model. This optimization cycle is iterated at the subsequent time step, generating a new problem solution as the time horizon shifts forward by one step. The scale of the MILP problem within RHO is significantly reduced compared to FHO, rendering RHO well-suited for real-time controllers. Notably, RHO can incorporate real-time measurements at every time step to guide optimal decision-making, and its accuracy surpasses that of FHO due to its capacity to handle system uncertainties. Under deterministic conditions and with a sufficiently extended time horizon, RHO equates to the FHO method. The problem posed by (\ref{eq:objfunc_linearize}) is reformulated within the RHO context, as reflected in (\ref{eq:objfuncRHO}).}

\begin{maxi}[1]
	{x}
	{	f(x) = f_1(o_i^t)
		- \omega_1 f_2(u_P^t)
	}
	{}{}
	\breakObjective{- \omega_2 f_3(u_{SoC}^t) + \omega_3 f_4(\text{SoC}^{T})},
	\label{eq:objfuncRHO}
\end{maxi}
\begin{align}
	x & = [o_i^t,P_e^{E,t}, P_g^{G,t}, u_P^t, u_{SoC}^t]^\top, \\
	\label{eq:objfuncf1RHO}
	f_1(o_i^t)
	  & = {
	\sum_{t = 1}^{N_p}
	{\sum_{i = 1}^{n_L}{\hat{w}_i o_i^t}}},                    \\
	f_2(P_e^{E,t})
	\label{eq:objfuncf2RHO}
	  & = \sum_{t = 1}^{N_p}{\sum_{e = 1}^{n_E}{u_P^t}},       \\
	f_3(\text{SoC}^t)
	\label{eq:objfuncf3RHO}
	  & = \sum_{t = 1}^{N_p}{\sum_{l,m}{u_{SoC}^t}},           \\
	f_4(\text{SoC}^{T})
	\label{eq:objfuncf4RHO}
	  & = \sum_{e = 1}^{n_E
	}{\alpha_e \text{SoC}_e^{E,T}},
\end{align}
where $N_p$ is the horizon length.

\section{Case Studies}
\label{sec: casestudies}
\noindent The RHO method is utilized for managing the ship's power system in situations with high ramp-rate loads. In this section, a comparison is provided between the RHO and FHO methods to demonstrate the effectiveness of the proposed RHO approach.

\subsection{System Description}

\begin{figure*}[t]
	\centering
	\includegraphics[width=0.7\linewidth]{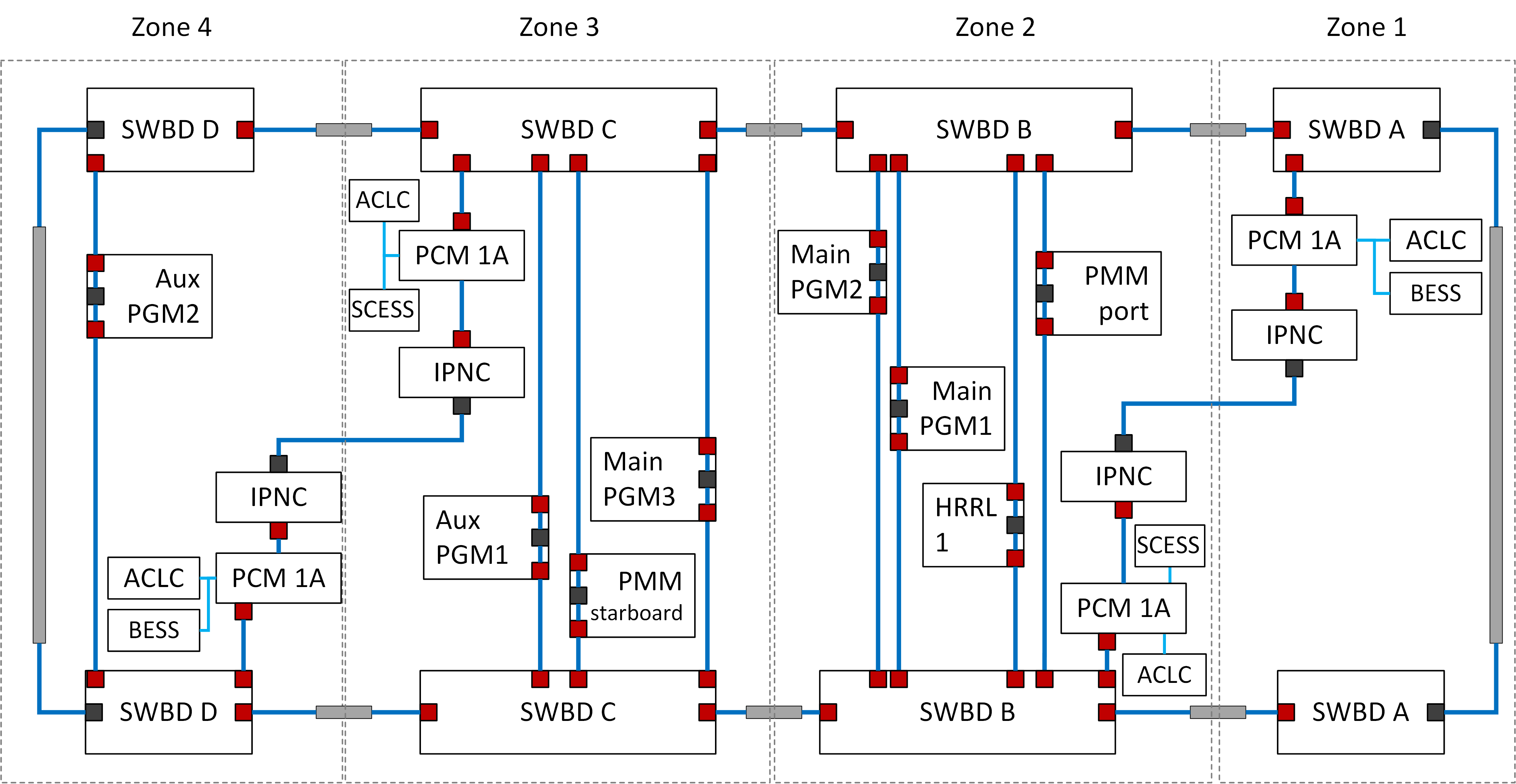}
	\caption{The four-zone MVAC ship power system.}
	\label{fig:MVAC_Notional}
\end{figure*}

\noindent \rev{As depicted in Fig.~\ref{fig:MVAC_Notional}, a four-zone SPS model is utilized for evaluating the proposed EMS. The shipboard power system has a power rating of 100 MW. Each zone consists of several modules, including the power conversion module (PCM-1A), propulsion motor module (PMM), integrated power node center (IPNC), power generation module (PGM), and energy storage module (ESM).} The tested system consists of three generators with a total capacity of 45~MW, two units of 10MW-BESS, and two units of 10MW-SCESS. The energy levels of BESS and SCESS are provided in Table~\ref{TB:systemparameters}, and the parameters of the RHO algorithm are given in Table~\ref{TB:controlparameters}. \rev{Additional details about this MVAC shipboard power system can be accessed in \cite{esrdc1270}.}

\begin{table}[t]
	\centering
	\renewcommand{\arraystretch}{1.3}
	\caption{\rev{Ship power system's characteristics}}
	\label{TB:systemparameters}
	\begin{tabularx}{3.3in}{c|c|c}
		\toprule
		Symbol                  & Parameter                        & Value         \\ 
		\midrule
		$n_E$                   & Number of ESS                    & 4             \\
		$P_E^{\max}$            & Maximum power of ESS             & $10$~MW       \\
		$P_E^{\min}$            & Minimum power of ESS             & $-10$~MW      \\
		$r_g^{\max,\min}$       & Generator's ramp-rate limitation & $\pm1$~MW/s   \\
		$r_{BESS}^{\max,\min}$  & BESS's ramp-rate limitation      & $\pm5$~MW/s   \\
		$r_{SCESS}^{\max,\min}$ & SCESS's ramp-rate limitation     & $\pm100$~MW/s \\
		$E_{BESS}$              & Energy capacity of BESS          & $1000$~MJ     \\
		$E_{SCESS}$             & Energy capacity of SCESS         & $200$~MJ      \\

		\bottomrule
	\end{tabularx}
\end{table}

\begin{table}[t]
	\renewcommand{\arraystretch}{1.3}
	\caption{\rev{Parameters of the RHO algorithm}}
	\label{TB:controlparameters}
	\centering
	\begin{tabularx}{2.5in}{c|c|c}
		\toprule
		Symbol     & Parameter                       & Value \\ 
		\midrule
		$\Delta t$ & \rev{Control sampling interval} & 0.5~s \\
		$N_p$      & Horizon length                  & 60    \\
		$T$        & Mission time                    & 600~s \\

		\bottomrule
	\end{tabularx}
\end{table}

\subsection{Gradient descent Algorithm to Optimize Objective Weights}
\label{sec: gradient decent}
\begin{figure}[t]
	\centering
	\includegraphics[width=0.9\linewidth]{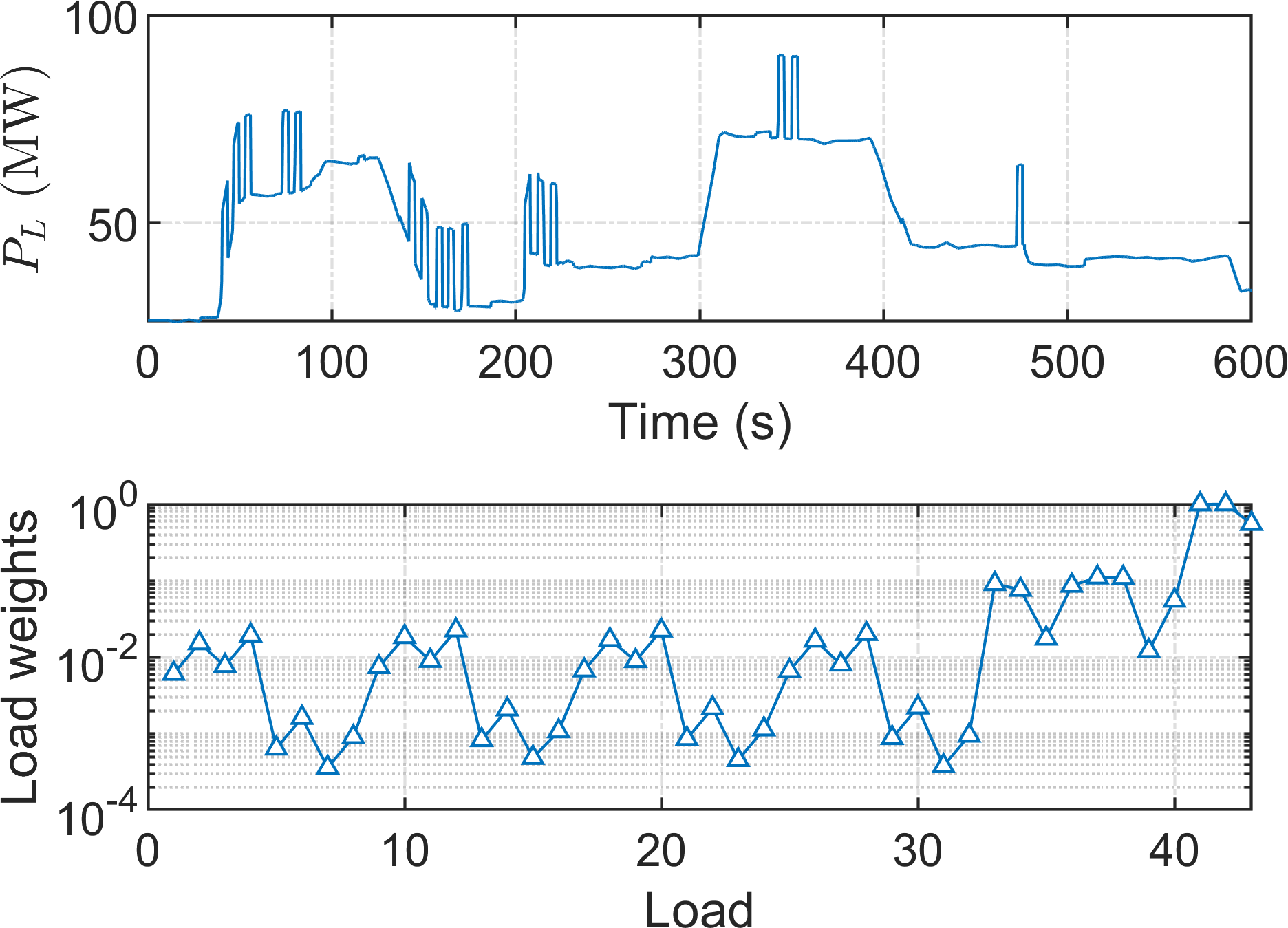}
	\caption{Total load profile and weights of 43 loads.}
	\label{fig:Loaddata}
\end{figure}
\begin{figure}[t]
	\centering
	\includegraphics[width=1.0\linewidth]{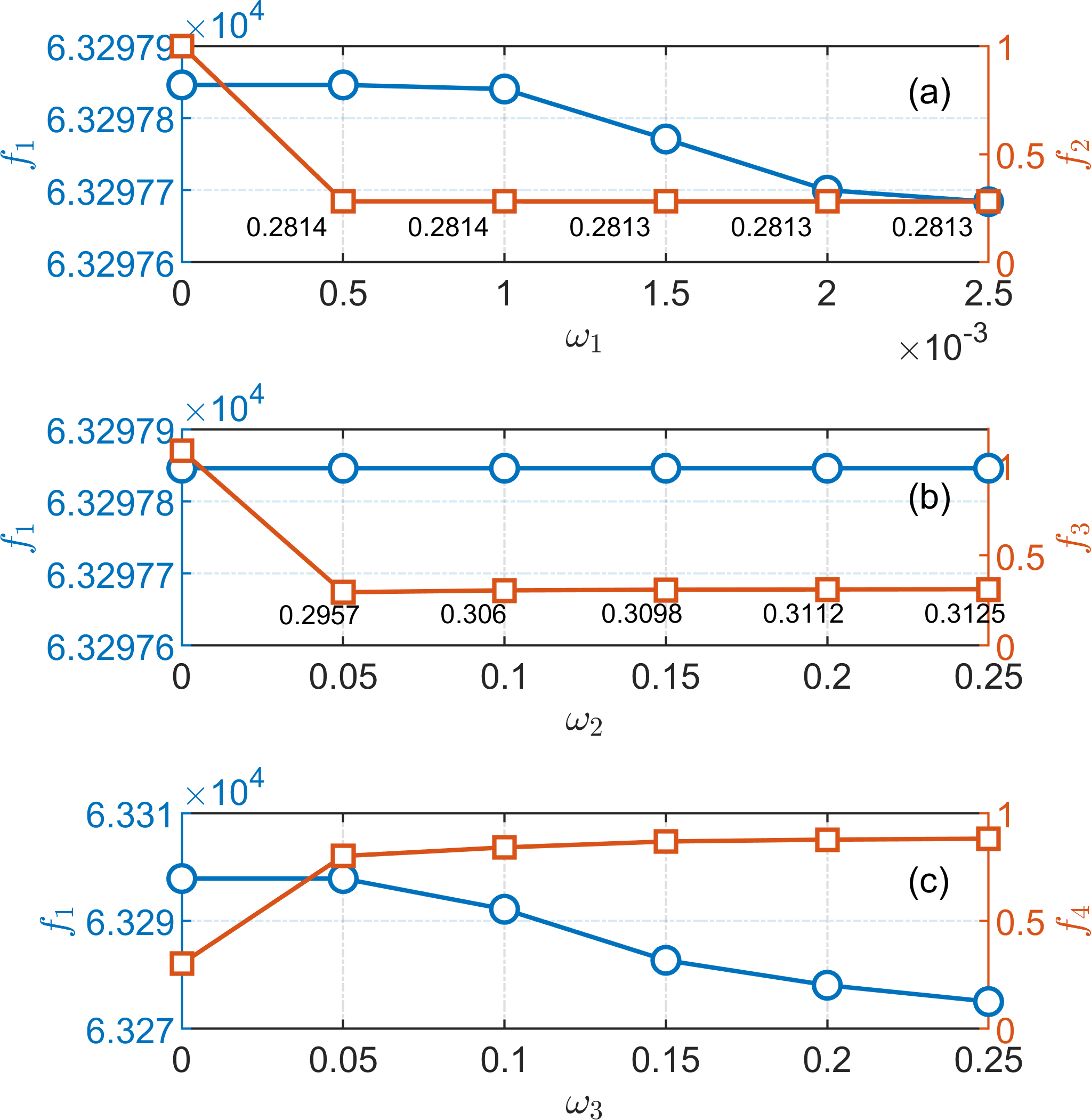}
	\caption{Effect of secondary objectives on load operability (high ramp-rate load profile.)}
	\label{fig:effectkgains}
\end{figure}
\begin{figure}[t]
	\centering
	\includegraphics[width=1.0\linewidth]{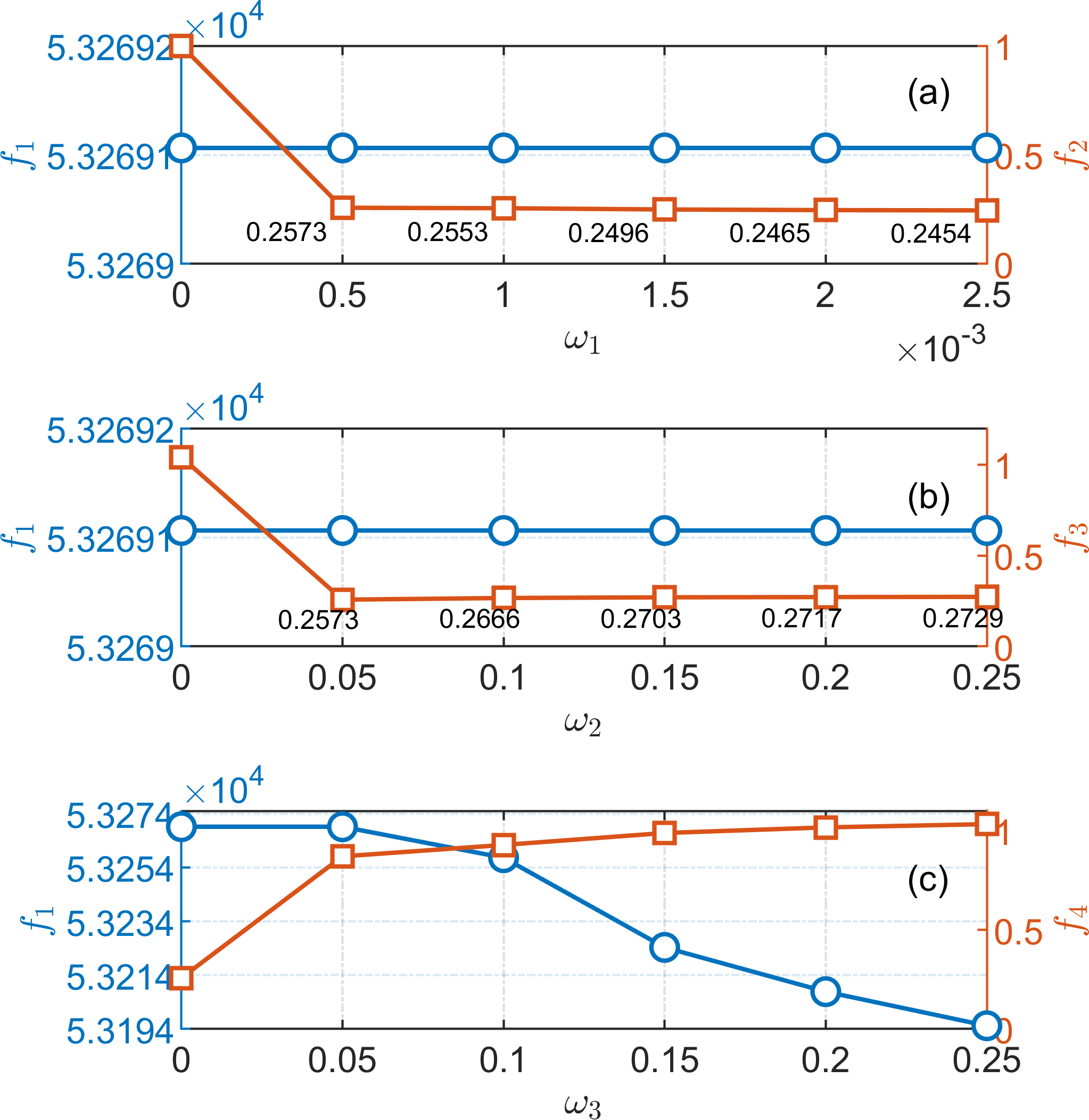}
	\caption{Effect of secondary objectives on load operability (baseline load profile.)}
	\label{fig:effectkgains42}
\end{figure}

\noindent The effects of constant weights, $\omega = [\omega_1, \omega_2, \omega_3]$, on load operability (\ref{eq:objfuncf1}) are examined in this section \revst{for both baseline and high ramp-rate load profiles}. Fig.~\ref{fig:effectkgains} shows the impacts of such weights on the main objective \revst{in case of the high ramp-rate load profile.} \rev{The profiles and weights of the 43 loads in the system are depicted in} Fig.~\ref{fig:Loaddata}. The increases of $\omega_1$ and $\omega_3$ result in reduced load operability. A high value of $\omega_1$ prevents ESS actions; a high value of $\omega_3$ makes ESS charge frequently to maximize SoC level, resulting in shedding more loads. On the other hand, constant weight $\omega_2$ slightly impacts load operability as it tries to balance the SoC level among ESS. \revst{The analysis of secondary objectives related to load operability also encompasses the baseline load profile, which represents a different mode of SPS without high ramp-rate loads, as illustrated in Fig.~\ref{fig:effectkgains42}. The findings align closely with those observed in the case of the high ramp-rate load profile with $f_3$ and $f_4$, while revealing a distinct impact of $f_2$. The increase in $\omega_1$ has only a marginal effect on load operability. This is attributed to the absence of high ramp-rate loads in the baseline profile, resulting in limited involvement of SCESS and consequently a smaller impact of the total absolute ESS power on the load operability, as described in (\ref{eq:objfuncf2}).}

From the above observation, it can be seen that function $\bar{f}$ in (\ref{eq:GDfunc}) is convex. Gradient descent (GD) algorithm in \textbf{Algorithm}~1 is used to find constant weights $\omega_i$. GD stop iterations when the absolute error of function $\bar{f}$ is smaller than a pre-defined value $\epsilon = 10^{-4}$. \revst{Optimal weights found by GD algorithm for high ramp-rate load profile are $\omega = [0.0056; 0.0321; 0.0541]$ and for baseline load profile are $\omega = [0.0059; 0.0333; 0.0566]$.}

\begin{algorithm}[ht]
	\begin{algorithmic}[1]
		\STATE $i \leftarrow 0$, initialize $\omega = [\omega_1; \omega_2; \omega_3]$
		\REPEAT
		\STATE Calculate $\bar{f}$ based on $\omega_1$, $\omega_2$, and $\omega_3$:
		$\bar{f}(i) \leftarrow \bar{f}|\omega$;

		\STATE $g(i) \leftarrow {\big[{\bar{f}(i) - \bar{f}(i-1)}\big] / \big[{\omega(i) - \omega(i-1)}\big]}$;


		\STATE	$\omega(i+1) \leftarrow \omega(i) - \gamma g(i);$
		\STATE $i \leftarrow i+1$
		\UNTIL{$|\bar{f}(i+1) - \bar{f}(i)| < \epsilon$}
		\RETURN $\omega$;
	\end{algorithmic}
	\label{alg:GDweights}
	\caption{Optimizing weights based on GD algorithm.}
\end{algorithm}

\begin{align}
	\label{eq:GDfunc}
	\bar{f} = -\bar{f}_1 + \bar{f}_2 + \bar{f}_3 - \bar{f}_4
\end{align}
where $\bar{f_i}$ is the normalized function of $f_i$.

\begin{figure}[t]
	\centering
	\includegraphics[width=0.9\linewidth]{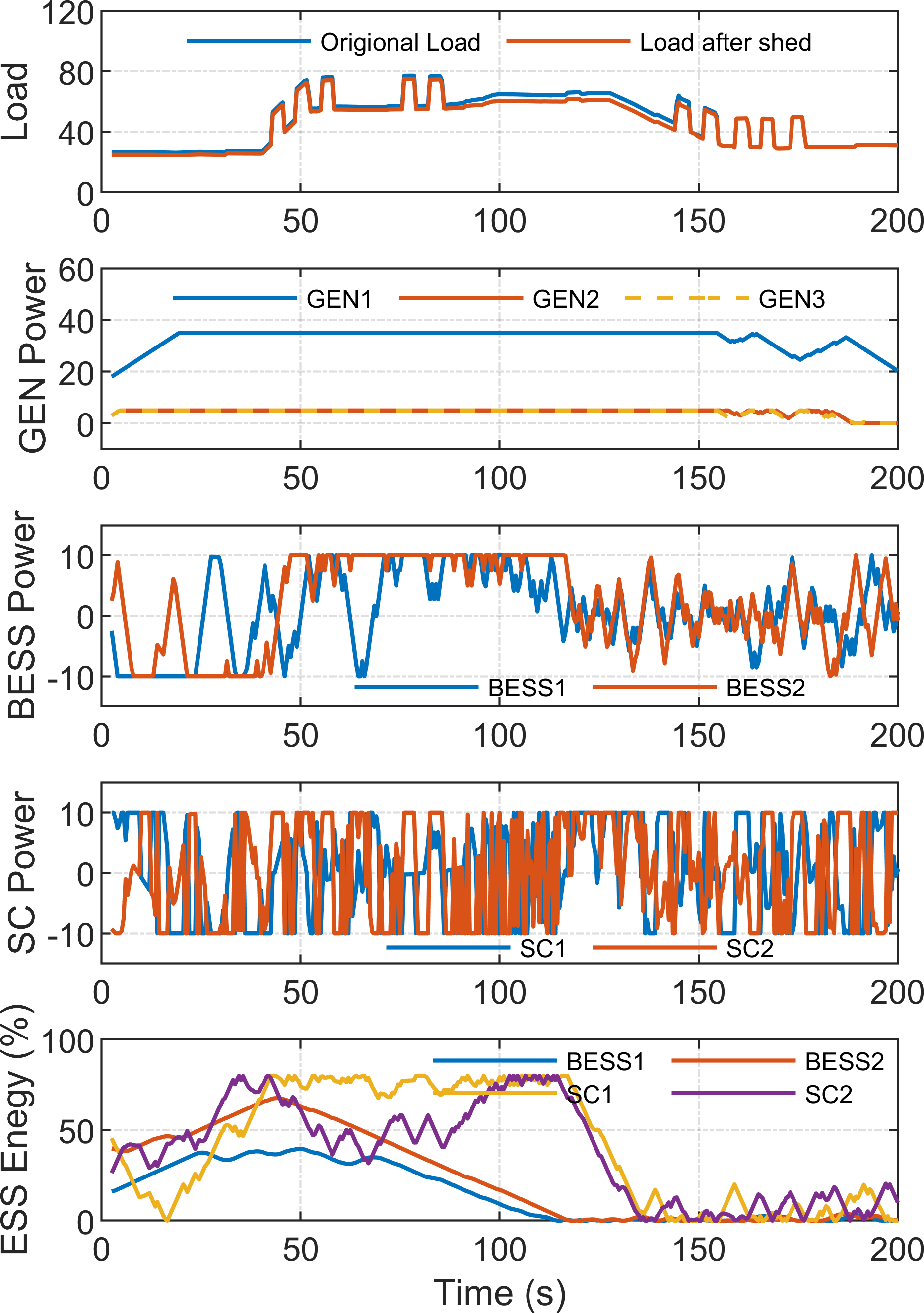}
	\caption{Optimization problem without secondary objectives ($f_1, f_2,$ and $f_3$.)}
	\label{fig:LSwoGD}
\end{figure}

\begin{figure}[t]
	\centering
	\includegraphics[width=0.9\linewidth]{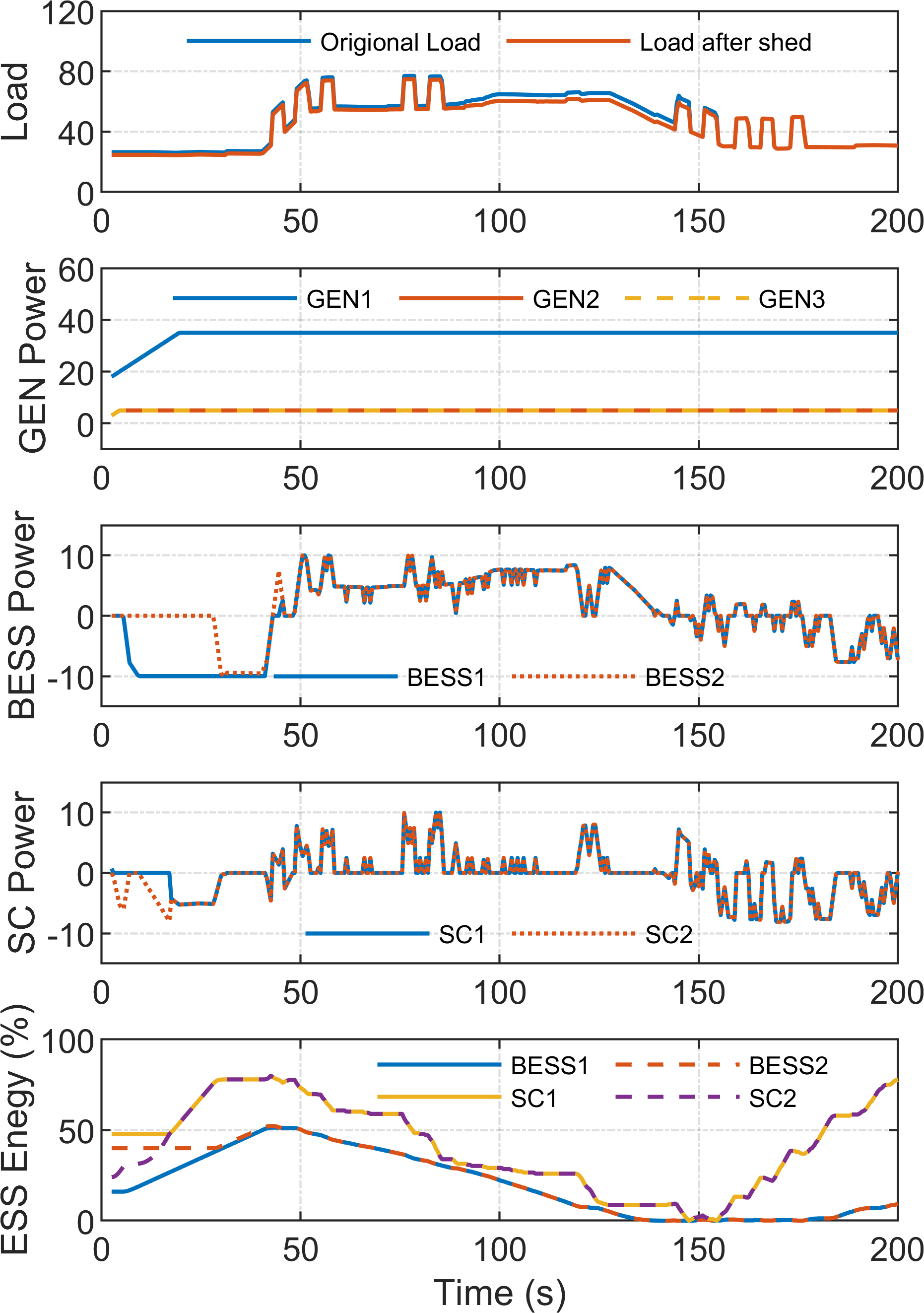}
	\caption{Optimization problem with optimal weights of secondary objectives.}
	\label{fig:LSwGD}
\end{figure}

Fig.~\ref{fig:LSwoGD} shows the optimization results of the ship power system without considering secondary objective terms of $f_1, f_2,$ and $f_3$. It can be seen in Fig.~\ref{fig:LSwoGD}(a) that loads are mainly disconnected between 80~s to 145~s. Several issues, such as circulating power among ESS and low SoC levels of ESS at the end of the optimization window, might cause ESS to fail in serving loads in the next mission. In addition, the SoC levels are different among ESS. Fig.~\ref{fig:LSwGD} shows the optimization results considering secondary objectives with optimal constant weights. It can be seen that the above issues are addressed. Initial SoC levels of four ESS are different; however, they are equal after 30~s due to optimally scheduling ESS powers. The SoC levels are maximized at every control sampling interval. In addition, since the SCESS are prioritized, their SoC levels are higher than those of BESS.
\revst{Fig.~\ref{fig:LSwGD}(c and d) also illustrate the optimal charge/discharge behaviors of BESS and SCESS in response to the load profile. BESS is charged during periods of surplus power (e.g., 0 - 40 s) and discharged during power shortages (e.g., 40 - 140 s). In contrast, SCESS primarily responds to high ramp-rate loads (e.g., 40 - 50, 140 - 180 s) to assist generators in serving loads, given the limited ramp rate of BESS (5 MW/s compared to 100 MW/s.)}

\subsection{Performance Evaluation}

\subsubsection{Performance under High Ramp-Rate Load Mission}
\noindent \rev{The performance of the proposed RHO method is evaluated within the context of HRRL operation. In this operating state, the rate of change in load power might exceed the generation's ramping capacity. Employing ESS, particularly SCESS, can offer assistance to the SPS during such instances.}

\begin{figure}[t]
	\centering
	\subfigure[All loads]
	{
		\includegraphics[width=0.85\linewidth]{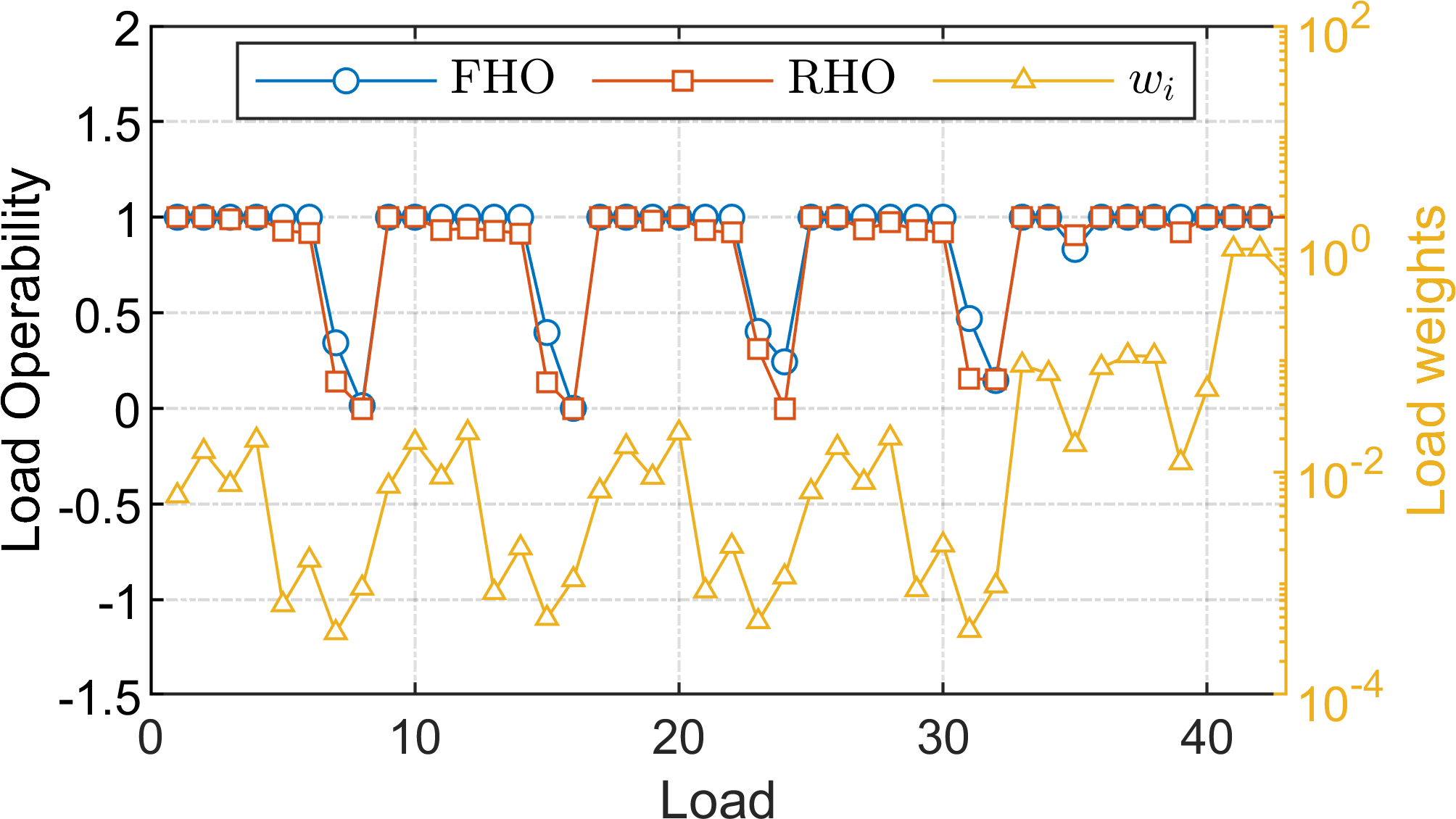}
	} \quad
	\subfigure[Loads 4 to 10]
	{
		\includegraphics[width=0.85\linewidth]{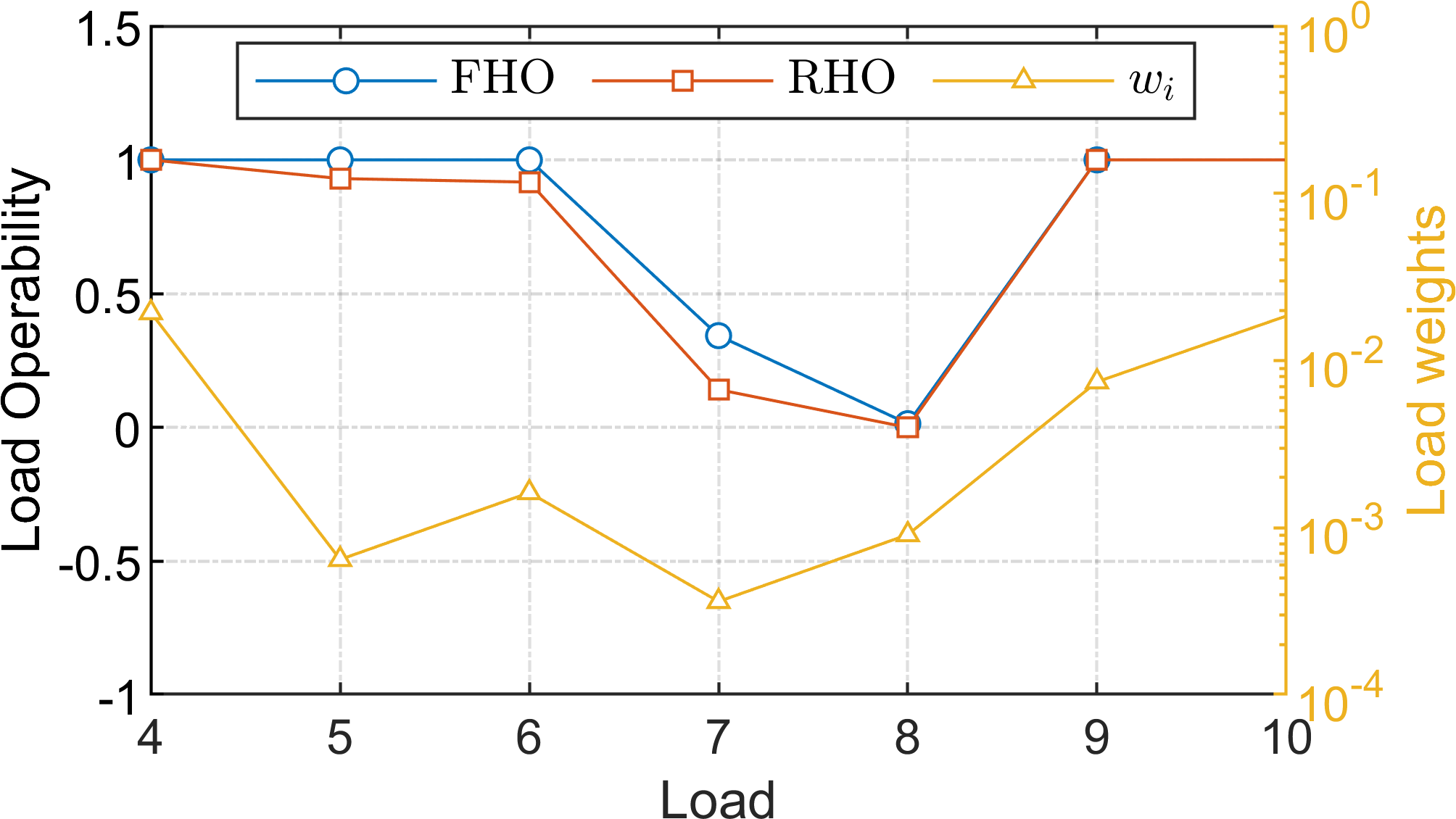}
	}
	\caption{Load operability comparison in HRRL mission.}
	\label{fig:Operability}
\end{figure}

\begin{figure}[t]
	\centering
	\subfigure[Whole mission]
	{
		\includegraphics[width=0.85\linewidth]{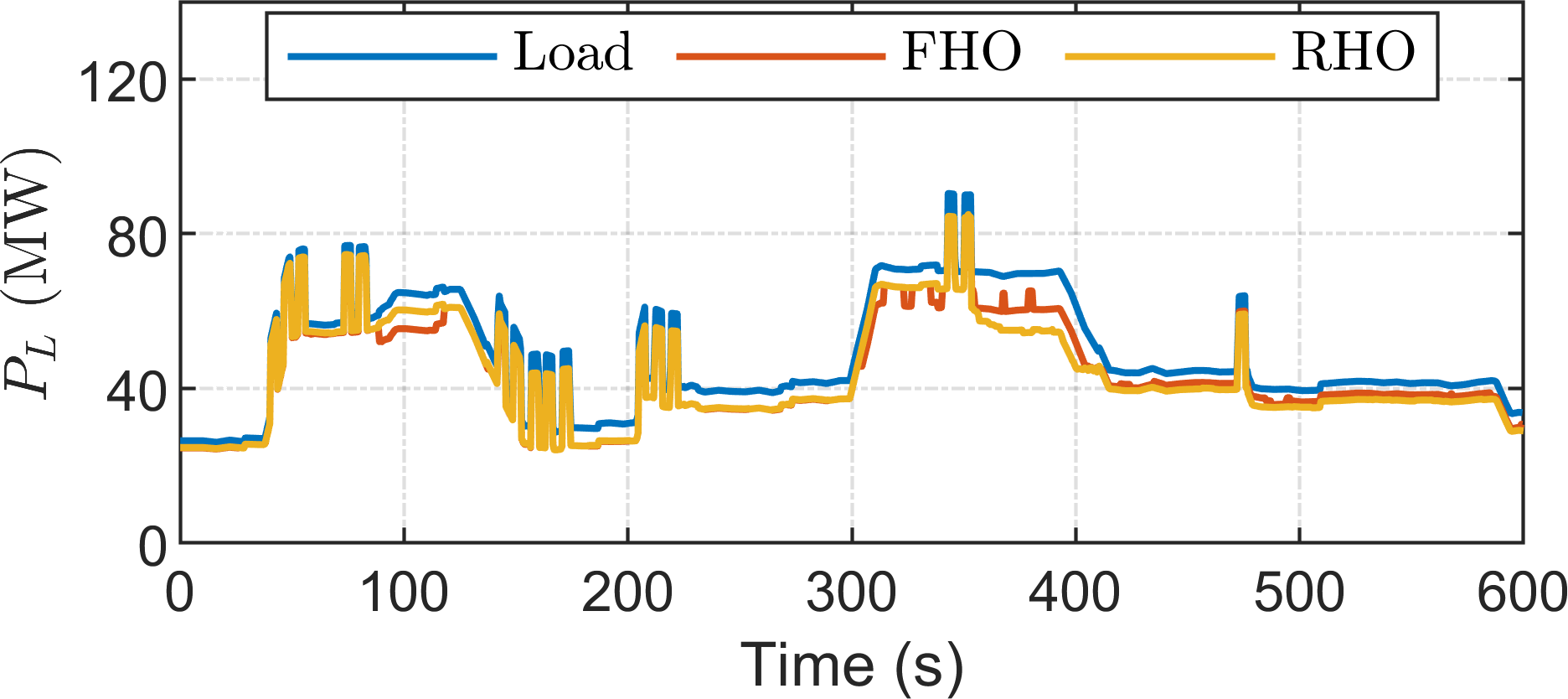}
	} \quad
	\subfigure[Zoom in]
	{
		\includegraphics[width=0.85\linewidth]{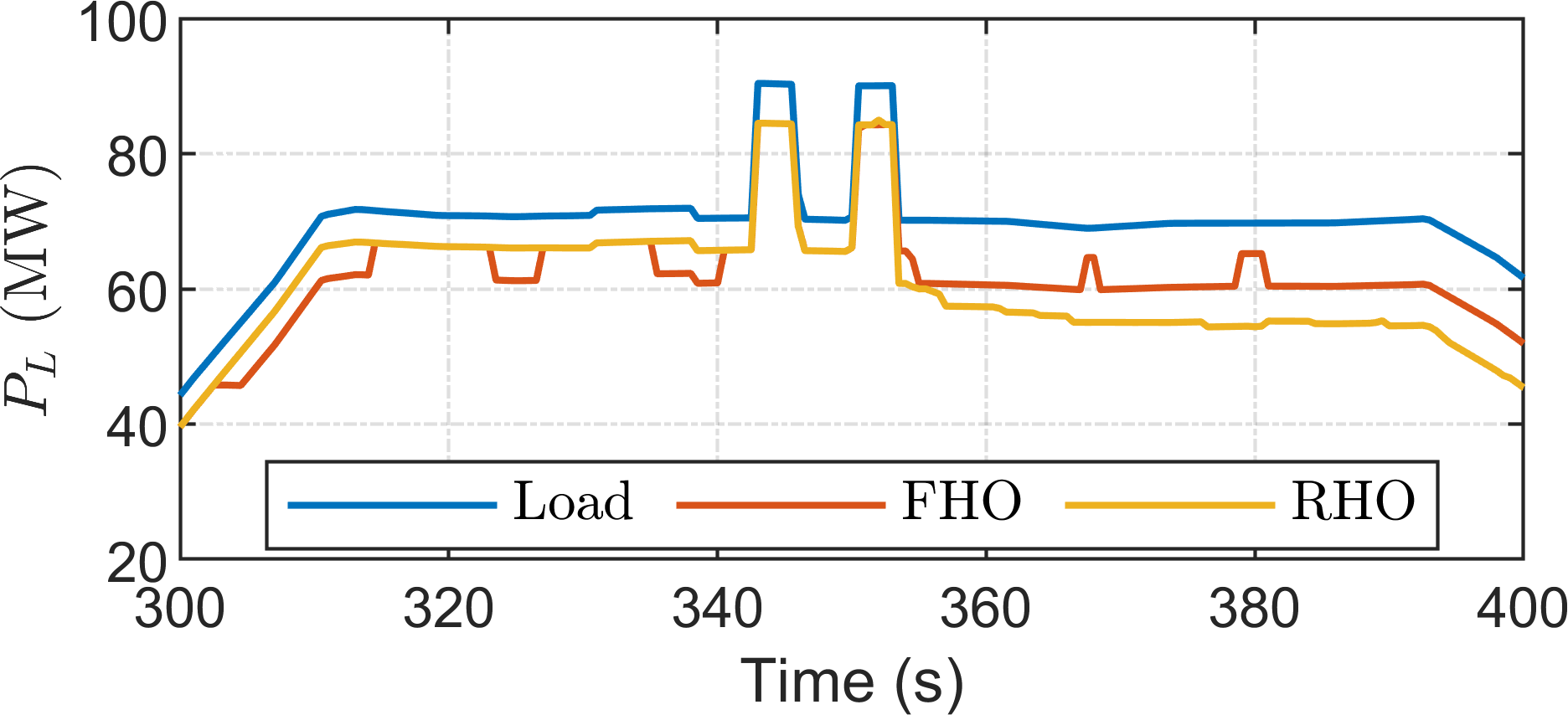}
	}
	\caption{\rev{Load profile and actual load served by FHO and RHO methods.}}
	\label{fig:Servedload}
\end{figure}

\rev{The HRRL loads, labeled as loads 41-43, bear the highest weight values, and consequently, demand highest priority. As showed in Fig.~\ref{fig:Operability}(a), load weights and the corresponding optimal load operability derived from two methods are depicted. Loads are curtailed when their operability falls below unity. Evidently, loads with lower weight values $\hat{w}_i$ are curtailed, whereas the HRRL loads with the greatest weight are prioritized for service, as illustrated in Fig.~\ref{fig:Operability}(b). Fig.~\ref{fig:Servedload}(a) presents the profiles of the actual served loads under both methodologies. For the FHO approach, a larger number of loads are curtailed before the 300~s mark to cater to the surge in load profile between 300 and 400~s, as highlighted in Fig.~\ref{fig:Servedload}(b). In comparison to FHO, RHO has an increased load curtailment period between 350 and 400~s, attributable to RHO's shorter optimization horizon relative to FHO. }The error of total load operability between the two methods given by (\ref{eq:deltaf}) indicates that the performance of the proposed RHO is close to FHO.

\begin{align}
	\label{eq:deltaf}
	\Delta f_1 = {\frac{{f_1^{FHO} - f_1^{RHO}}}{{f_1^{FHO}}}} = +0.05\%
\end{align}

\begin{figure}[t]
	\centering
	\includegraphics[width=0.9\linewidth]{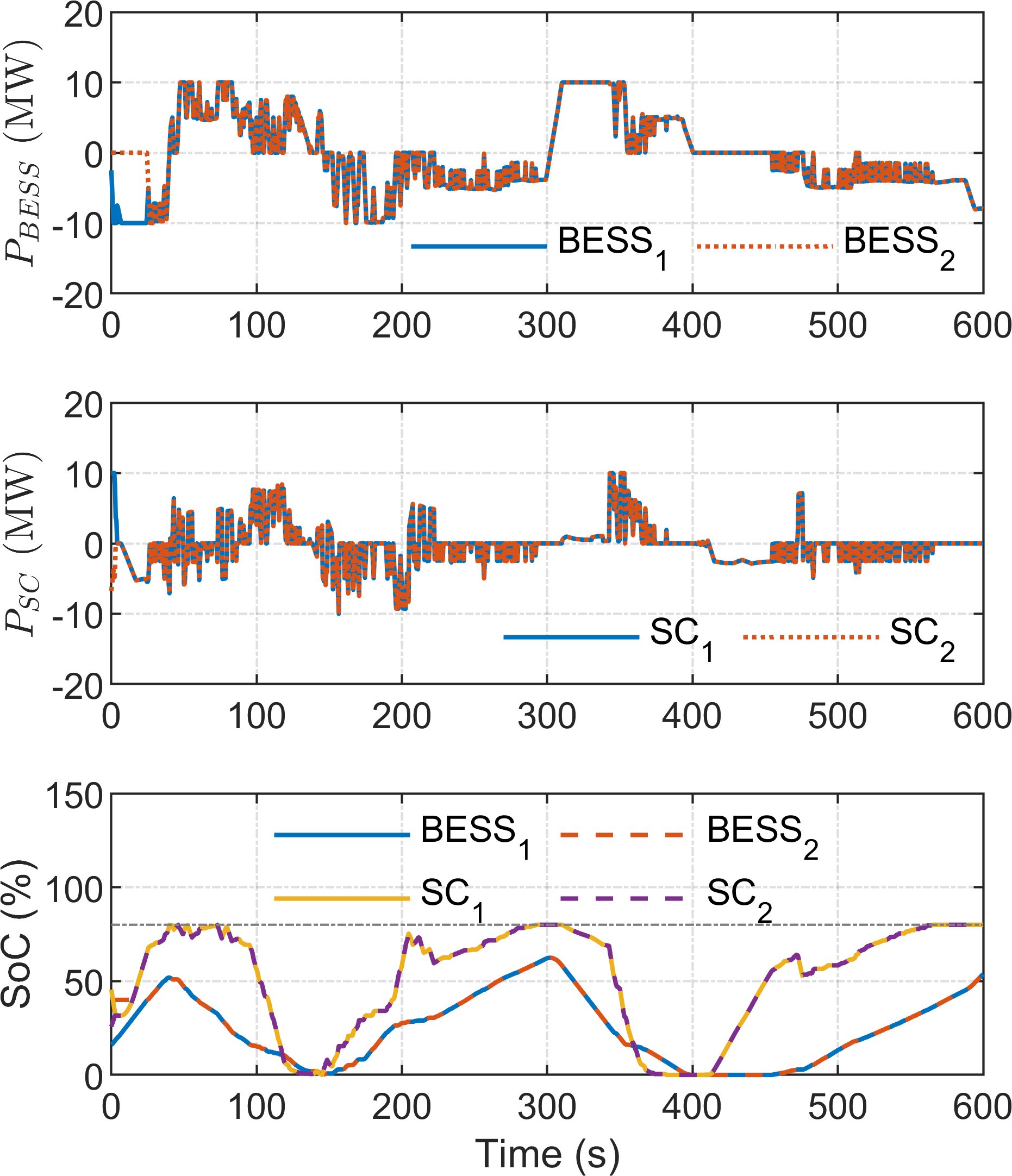}
	\caption{ESS power and SoC of the RHO approach.}
	\label{fig:RHO_ESS}
\end{figure}

\begin{figure}[t]
	\centering
	\includegraphics[width=0.9\linewidth]{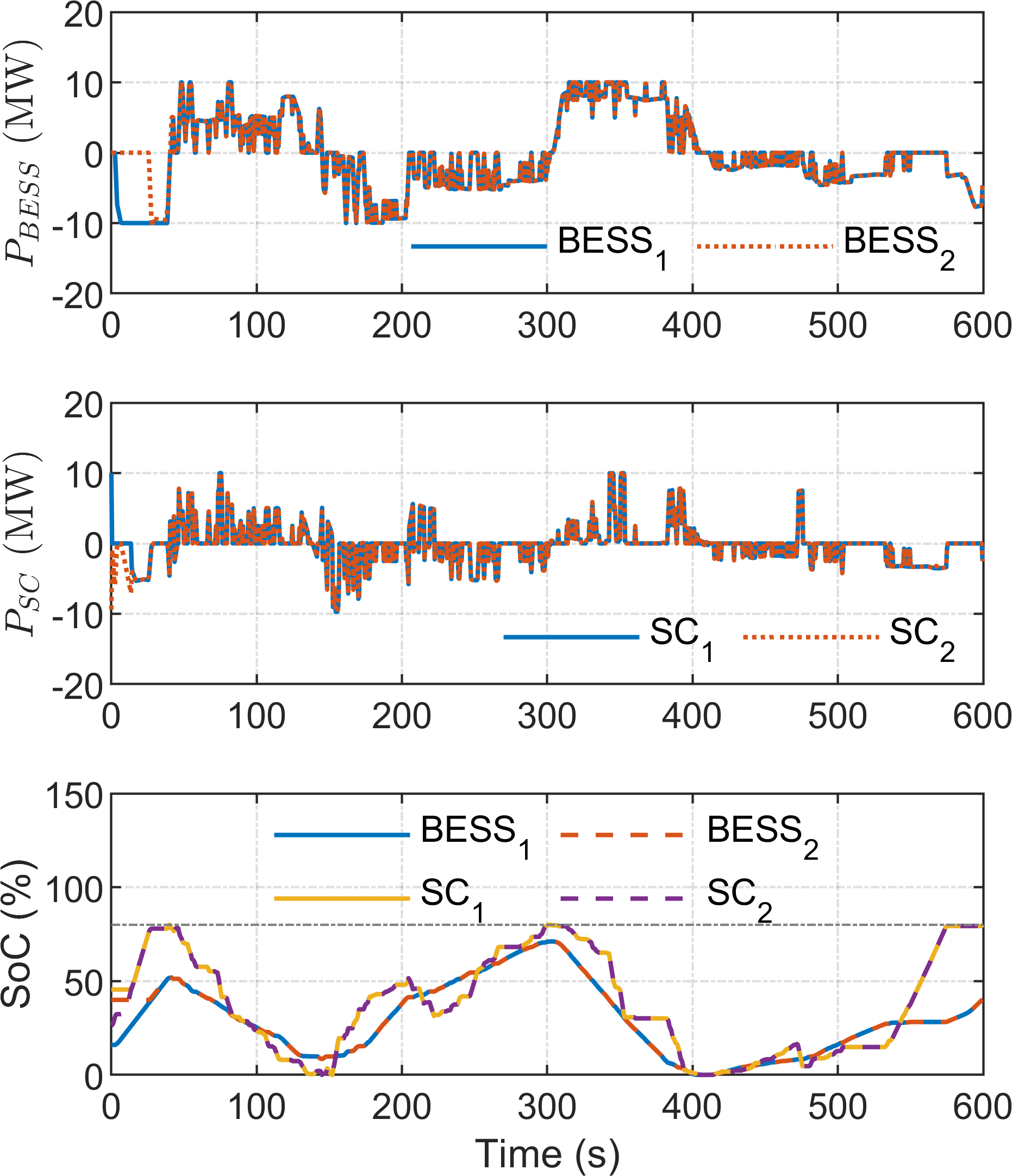}
	\caption{ESS power and energy when using FHO}
	\label{fig:FHO_ESS}
\end{figure}

\rev{The output power results of BESS and SCESS are illustrated in Figs.~\ref{fig:RHO_ESS} and \ref{fig:FHO_ESS}, respectively. The timeframe from 0~s to 50~s corresponds to a scenario where four ESS are actively charging due to an excess of generation compared to the load demand. Owing to variations in the original SoC levels among the ESS units, their charging rates differ based on SoC balance. Notably, after 50~s, there is a convergence in SoC levels between two BESS and two SCESS units, resulting in equivalent output powers for either two BESS or two SCESS units. It should be noted that the SoC of ESS units is limited to a maximum of 80\%. Within the framework of both methodologies, SCESS is favored for charging, leading to higher SoC levels at the conclusion of the optimization window compared to BESS. To this end, the performance of the proposed RHO technique aligns closely with that of the FHO approach.}

\subsubsection{Effect of Load's Weights on RHO Performance}
\begin{figure}[!tbp]
	\centering
	\includegraphics[width=0.95\linewidth]{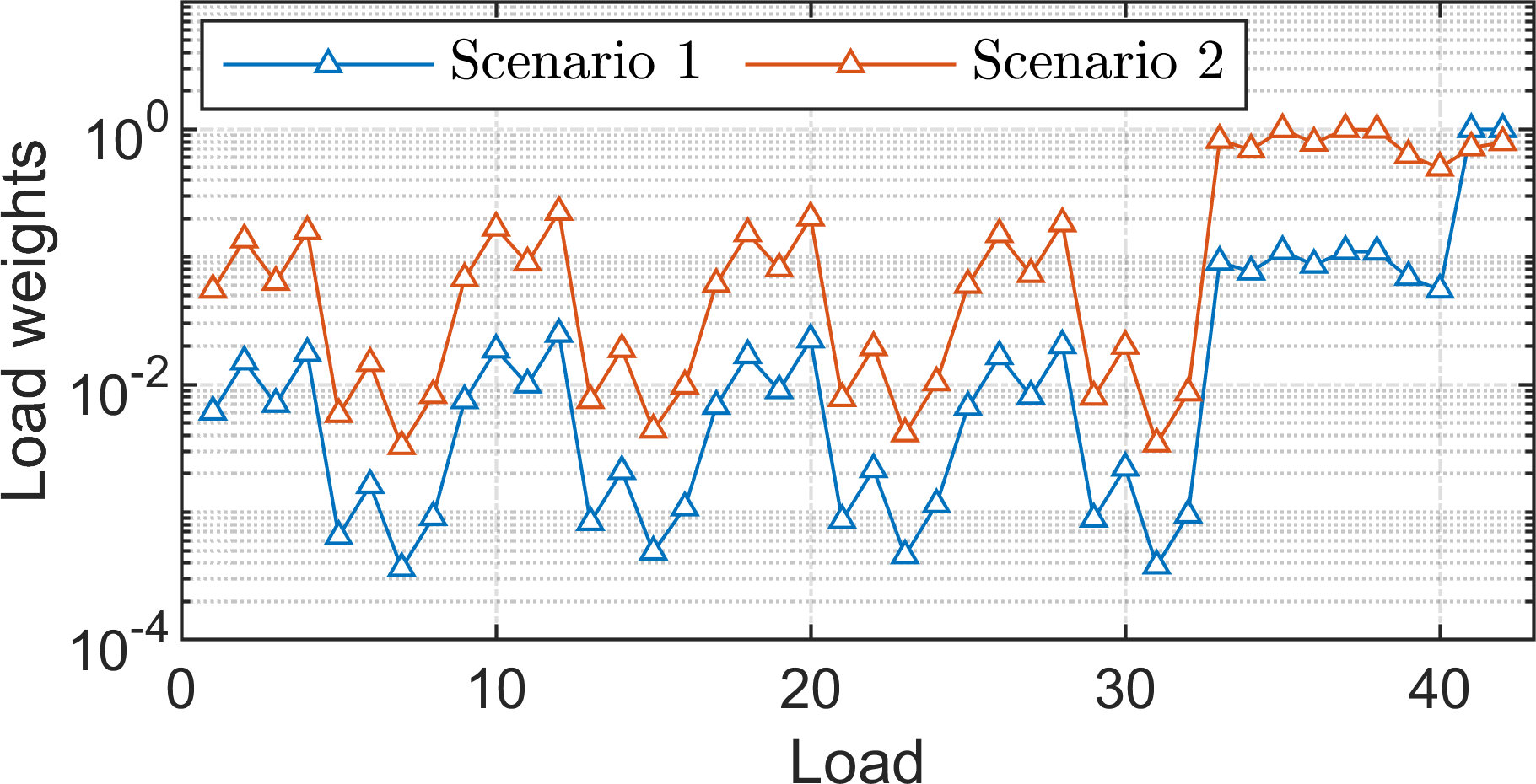}
	\caption{Weights of loads for different scenarios.}
	\label{fig:Lweights}
\end{figure}
\begin{figure}[!tbp]
	\centering
	\includegraphics[width=1.0\linewidth]{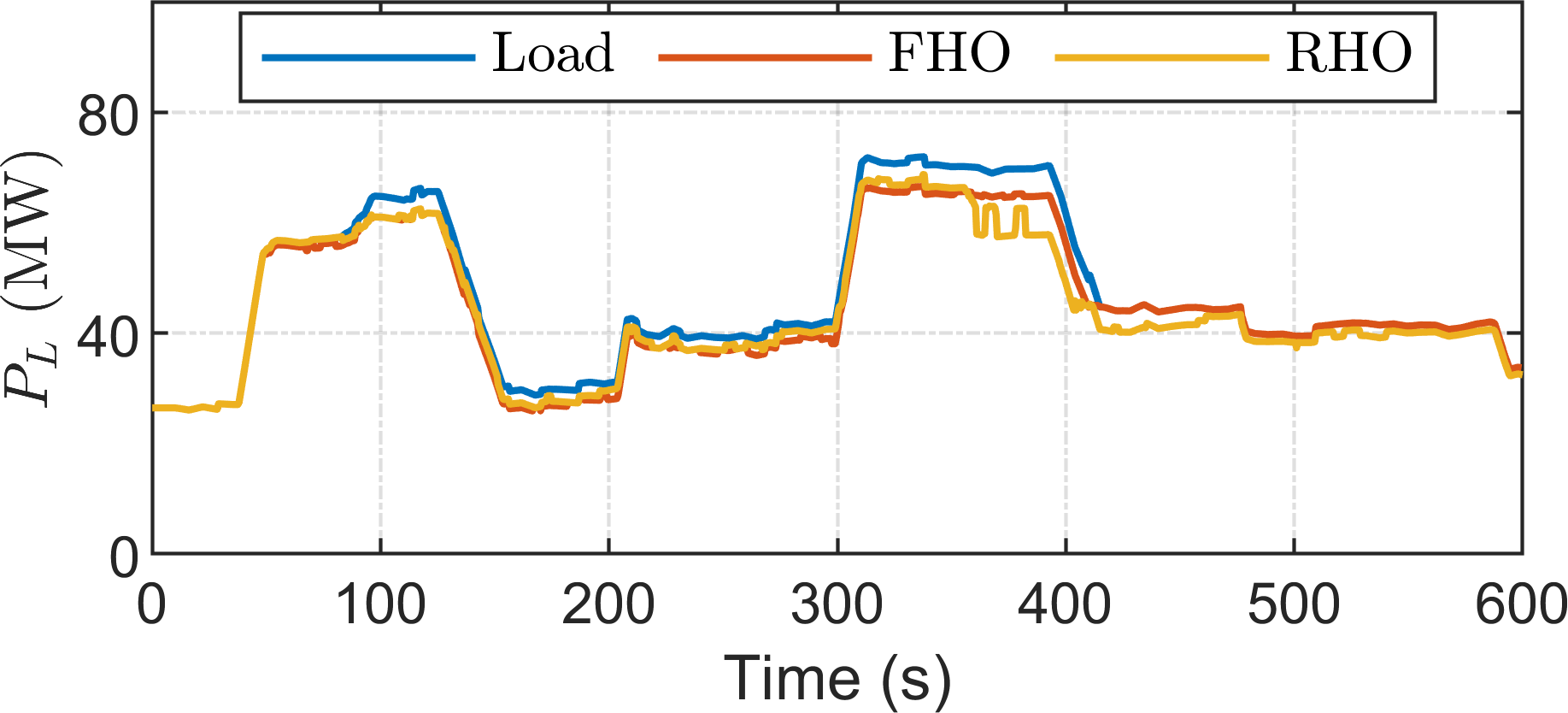}
	\caption{Served load in scenario 1.}
	\label{fig:LSbased}
\end{figure}
\begin{figure}[!tbp]
	\centering
	\includegraphics[width=1.0\linewidth]{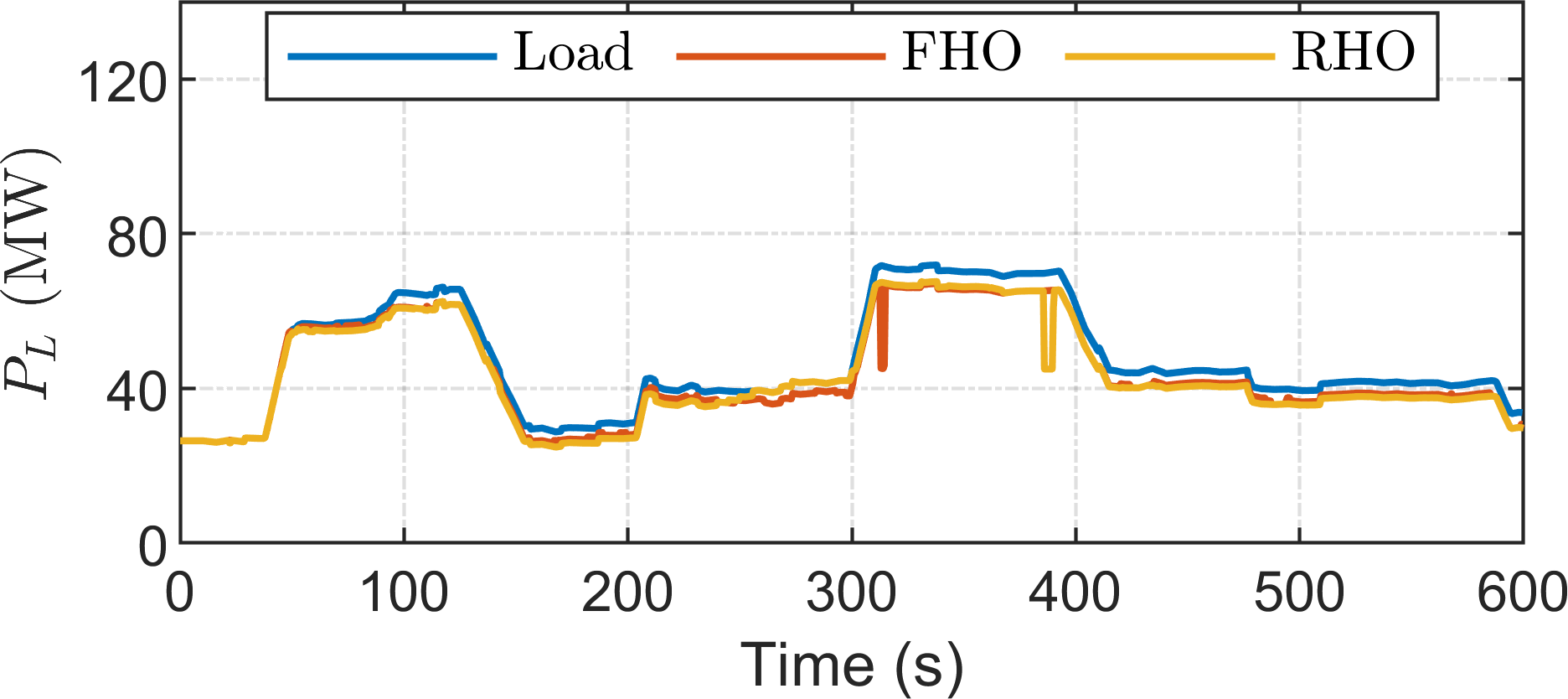}
	\caption{Served Load in scenario 2.}
	\label{fig:LSlowPMM}
\end{figure}
\begin{figure}[t]
	\centering
	\includegraphics[width=0.92\linewidth]{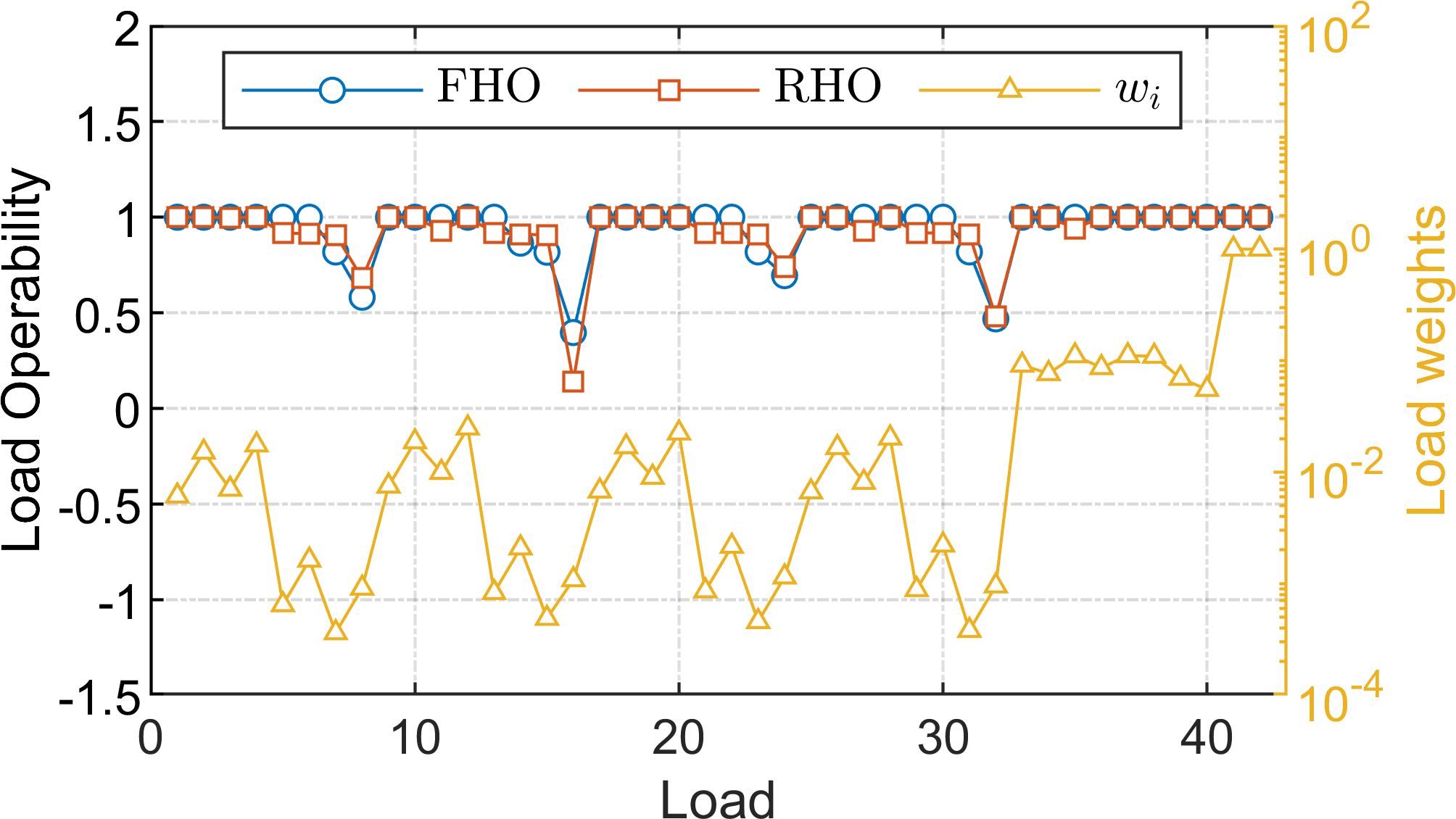}
	\caption{Compare load operability in scenario 1.}
	\label{fig:LSbasedOi}
\end{figure}
\begin{figure}[t]
	\centering
	\includegraphics[width=0.92\linewidth]{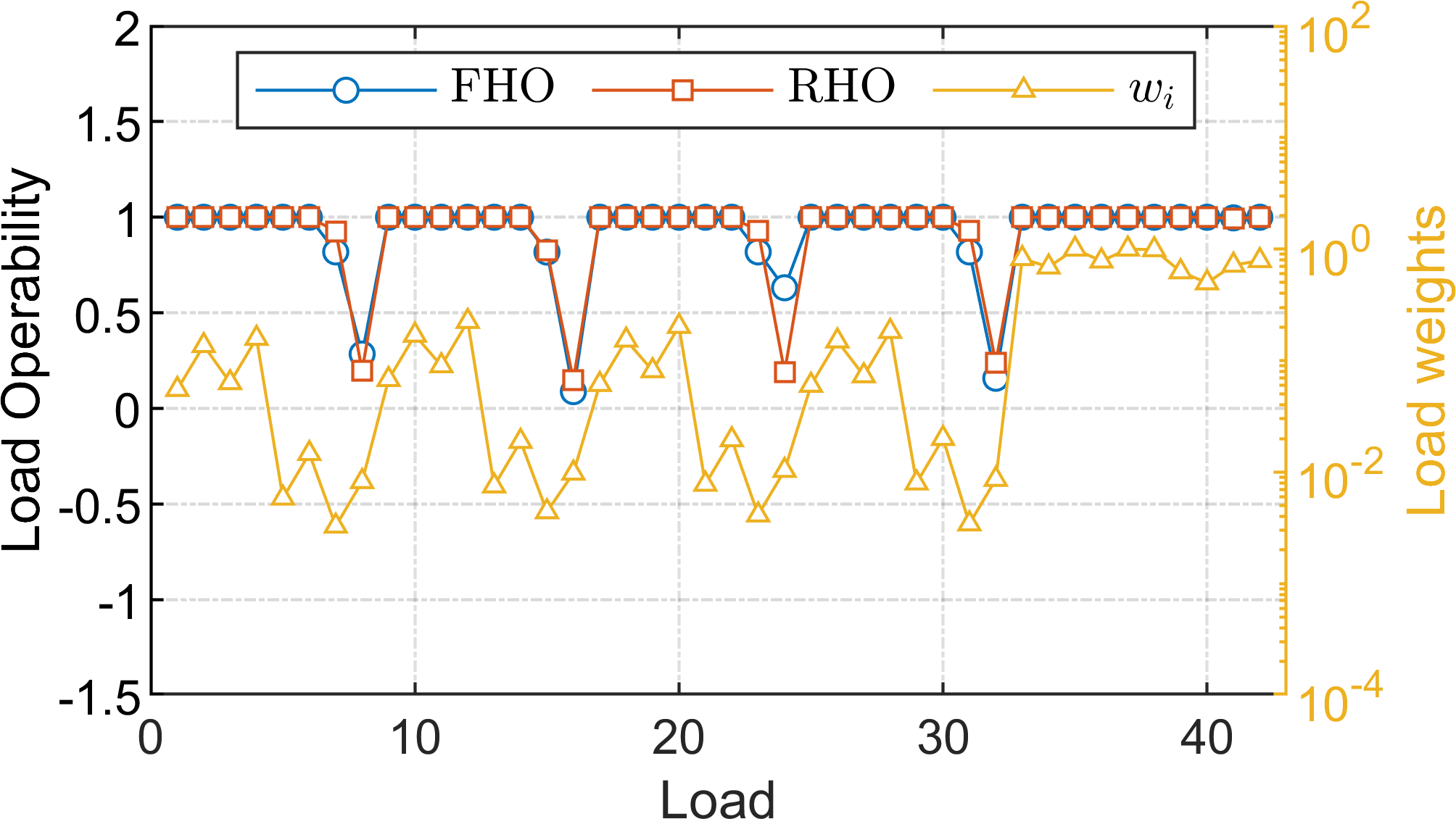}
	\caption{Compare load operability in scenario 2.}
	\label{fig:LSlowPMMOi}
\end{figure}

\noindent The ship's MVAC power system operates under various scenarios, such as peacetime cruise, sprint station, battle, and anchor \cite{esrdc1270}. The importance of loads varies with the operating conditions. A load might be vital in one scenario but not in another. For example, when the mission of the ship power system changes from cruise to battle, some loads that were previously nonvital become vital. The proposed RHO method is evaluated by two additional scenarios, as shown in Fig. \ref{fig:Lweights}.

Figs.~\ref{fig:LSbased} and \ref{fig:LSlowPMM} show a comparison of served loads between proposed RHO and FHO methods under scenarios 1 and 2, respectively. Loads are mainly shed from 50~s to 400~s, and only loads with low weight values are mainly shed, as shown in Figs.~\ref{fig:LSbasedOi} and \ref{fig:LSlowPMMOi}. In scenario 2, instead of shedding significant loads at 314~s (FHO), the RHO method sheds significant loads at 385~s. As the optimization window of the RHO method is much shorter than that of the FHO method, loads are still served at 314~s. All ESSs are almost fully discharged after serving loads from 314~s to 384~s; loads are then shed significantly at 385~s due to the shortage of power generation. The total load operability in Figs. \ref{fig:LSbasedOi} and \ref{fig:LSlowPMMOi} indicates which loads are mainly shed in two scenarios. \rev{In both cases, nonvital loads with lower weight values are curtailed, while essential loads with higher weight values are ensured supply. In summary, the RHO method achieves a performance akin to that of the FHO method.} \\

\subsection{Computational Comparison}
\begin{figure}[t]
	\centering
	\includegraphics[width=0.85\linewidth]{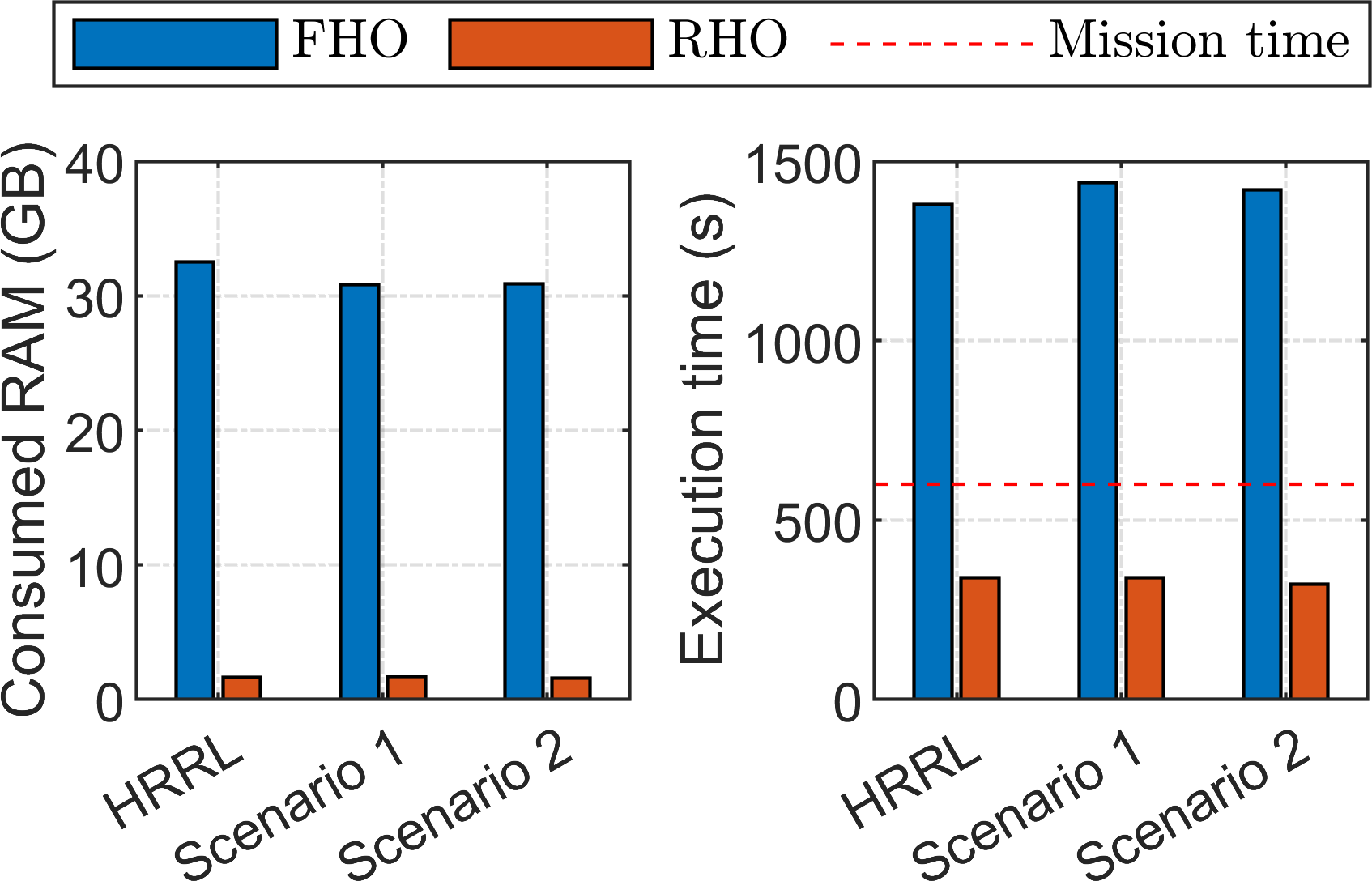}
	\caption{Computational contrast between FHO and RHO. The total execution time for the RHO approach comprises the cumulative time taken by all timesteps over a 600~s mission.}
	\label{fig:LScomputation}
\end{figure}
\begin{figure}[t]
	\centering
	\includegraphics[width=0.9\linewidth]{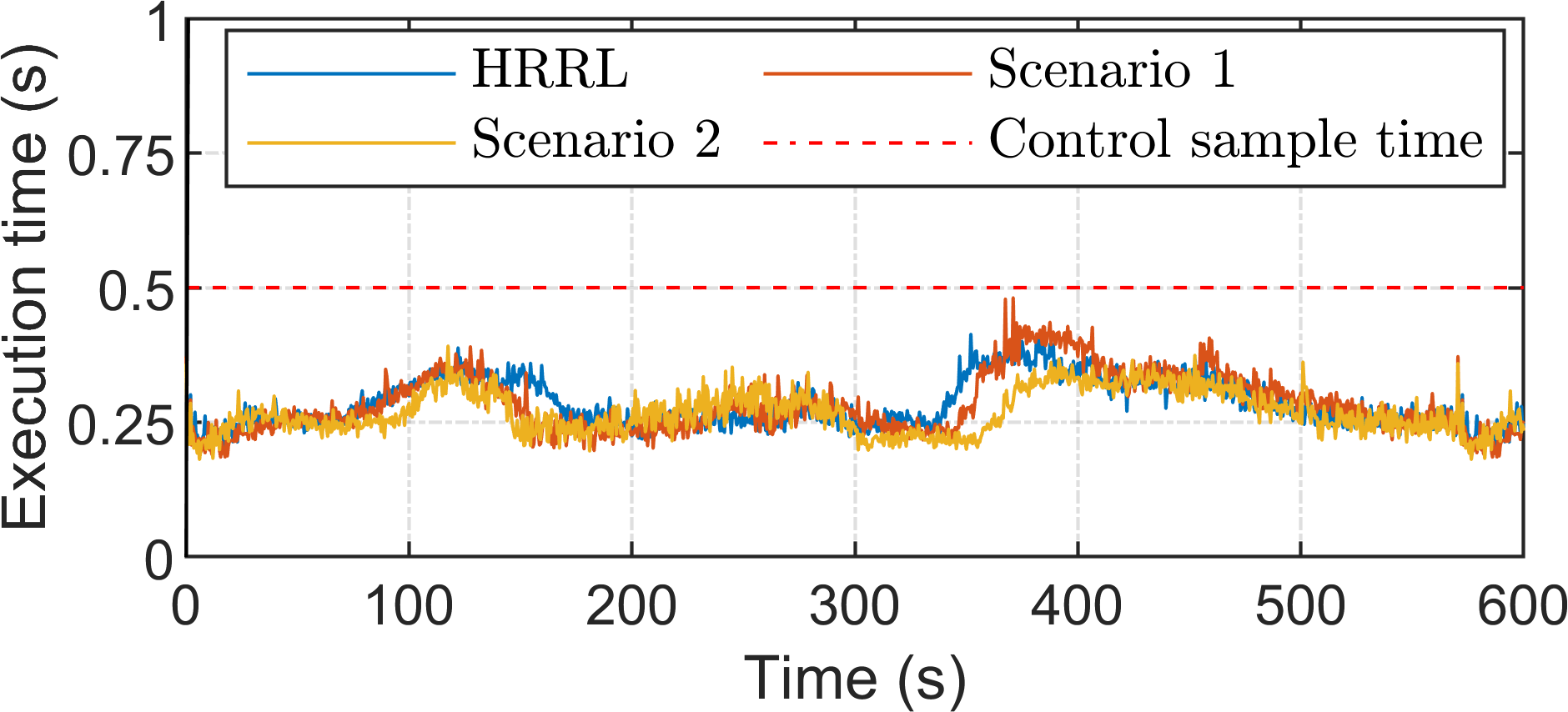}
	\caption{Execution time of the RHO method measured at each optimization step.}
	\label{fig:LSexecutetime}
\end{figure}

\noindent It was demonstrated in the above sections that the proposed methodology achieved comparable performance to the conventional optimization approach (FHO method). \rev{However, the true advantage of the RHO lies in its computational efficiency, which is demonstrated in this section.}

\rev{The simulations were conducted using an $\text{Intel}^{\circledR}$ $\text{Core}^{\text{TM}}$ i7-10700H processor running at 2.90~GHz and 64~GB RAM}. The \textit{intlinprog} function based on \textit{primal-simplex} algorithm from MATLAB is used to solve the MILP problem. \rev{Different scenarios were carried out to assess the computational efficiency of both approaches, such as HRRL, scenarios 1 and 2. In all scenarios, the total mission time is~600~s.}

\rev{A comprehensive evaluation between two methods in terms of computational performance is illustrated in Fig.~\ref{fig:LScomputation}. The FHO method necessitates around 31-33~GB RAM for solving the problem, while the RHO method operates efficiently with only 1.6-1.7~GB RAM. Furthermore, the FHO method takes around 1380-1440~s to complete the task, while the RHO method achieves the same outcome in under 400~s. These substantial reductions in both memory usage and execution time underscore the enhanced efficiency and practicality offered by the proposed RHO method.}

The optimal solution must be found before applying the control actions in order to ensure real-time applications. In the case of FHO, the execution time must be less than 600~s; in the case of RHO, the execution time at each step must be less than the control sample time of 0.5~s. It can be seen that the FHO method is not applicable for real-time implementation, as its execution time is two times higher than the required time of 600~s. \rev{In contrast, the proposed RHO method demonstrates its suitability for real-time implementation by achieving optimization within a time frame of less than 0.5~s, as illustrated in Fig. \ref{fig:LSexecutetime}. This further reinforces its practicality and efficiency for real-time applications. Moreover, when compared to the FHO method, the RHO method exhibits substantial advancements in computational performance. Specifically, the RHO method showcases a remarkable reduction in RAM usage, with a decrease of 95\%. Additionally, it achieves an execution time that is approximately four times faster than the FHO method. These improvements highlight the significant benefits of employing the RHO method for the EMS.}

\section{Conclusion}
\label{sec:conclusion}
A resilience-oriented EMS of SPS has been proposed in this paper. Its advantages lie in improving computational efficiency for real-time applications by adopting the RHO technique. In addition, the proposed method utilized different ramp-rate characteristics of ESS to maximize load operability, resulting in the enhancement of system resilience. Practical constraints of ESS, such as SoC balancing and ESS prioritization, were considered in the proposed approach. The gradient descent algorithm was proposed to optimally design the weights of multiple objectives, ensuring that the performance of the main objective is preserved. \rev{A comparative analysis between the RHO method and the FHO method was conducted, demonstrating comparable resilient outcomes. However, the implementation of the RHO method exhibited notable advantages in terms of computational efficiency, with significantly reduced RAM consumption (95\% lower) and faster execution time (four times faster). The current study assumes sufficient power cable capacity to facilitate power transfer among subsystems. In our future work, we plan to incorporate power line limitations implement on real-time platforms such as OPAL-RT or RTDS. This will enable a more comprehensive evaluation of the system's performance and enhance its practical applicability.}

\section*{Acknowledgment}

This material is based upon research supported by, or in part by, the U.S. Office of Naval Research under award number N00014-16-1-2956.


\bibliographystyle{elsarticle-num}
\bibliography{main}

\end{document}